\documentclass[twocolumn]{aastex63}

\newcommand{\beq}{\begin{equation}}
\newcommand{\eeq}{\end{equation}}

\usepackage{amsmath,amstext}
\usepackage[T1]{fontenc}
\usepackage{apjfonts} 
\usepackage[figure,figure*]{hypcap}
\usepackage[normalem]{ulem}             

\usepackage{graphicx}

\usepackage{color}

\renewcommand{\ion}[2]{#1\,{\sc #2}}

\submitjournal{ApJ}

\shorttitle{Multi-wavelength observations of AR12759. }
\shortauthors{Del Zanna, G. et al.}

\begin{document}

\title{Multi-wavelength observations by XSM, Hinode and SDO of an active region. Chemical abundances and temperatures}

\correspondingauthor{G. Del Zanna}
\email{gd232@cam.ac.uk}

\author[0000-0002-4125-0204]{G. Del Zanna}
\affiliation{DAMTP, Center for Mathematical Sciences, University of Cambridge, Wilberforce Road, Cambridge, CB3 0WA, UK}
\author[0000-0002-7020-2826]{B. Mondal}
\affiliation{Physical Research Laboratory, Navrangpura, Ahmedabad, Gujarat-380 009, India }
\affiliation{Indian Institute of Technology Gandhinagar, Palaj, Gandhinagar, Gujarat-382 355, India}
\author[0000-0002-3064-0610]{Yamini K. Rao}
\affiliation{DAMTP, Centre for Mathematical Sciences, University of Cambridge, Wilberforce Road, Cambridge CB3 0WA, UK}
\author[0000-0003-3431-6110]{N. P. S. Mithun}
\affiliation{Physical Research Laboratory, Navrangpura, Ahmedabad, Gujarat-380 009, India }
\author[0000-0002-2050-0913]{S. V. Vadawale}
\affiliation{Physical Research Laboratory, Navrangpura, Ahmedabad, Gujarat-380 009, India }
\author[0000-0002-6903-6832]{K. K. Reeves}
\affiliation{Harvard-Smithsonian Center for Astrophysics, 60 Garden St, MS 58, Cambridge, MA, 20138}
\author[0000-0002-6418-7914]{H. E. Mason}
\affiliation{DAMTP, Centre for Mathematical Sciences, University of Cambridge, Wilberforce Road, Cambridge CB3 0WA, UK}
\author[0000-0002-4781-5798]{A. Sarkar}
\affiliation{Physical Research Laboratory, Navrangpura, Ahmedabad, Gujarat-380 009, India }
\author[0000-0003-2504-2576]{P. Janardhan}
\affiliation{Physical Research Laboratory, Navrangpura, Ahmedabad, Gujarat-380 009, India }
\author[0000-0003-1693-453X]{Anil Bhardwaj}
\affiliation{Physical Research Laboratory, Navrangpura, Ahmedabad, Gujarat-380 009, India }

\begin{abstract}
We have reviewed the first year of observations of the Solar X-ray Monitor (XSM) onboard Chandrayaan-2, and the available multi-wavelength observations to complement the XSM data, focusing on Solar Dynamics Observatory AIA and Hinode XRT, EIS observations. XSM has provided disk-integrated solar
spectra in the 1--15 keV energy range, observing a large number of microflares. 
We present an analysis of multi-wavelength observations of AR 12759 during its disk crossing.
We use a new radiometric calibration of EIS  to find that the quiescent AR core emission during
its disk crossing has a  distribution of temperatures and chemical abundances that does not change
significantly over time.
An analysis of the XSM spectra confirms the EIS results, and shows that the low
First Ionization Potential (FIP) elements are enhanced, compared to their photospheric values.
The frequent microflares produced by the AR did not affect the abundances of the quiescent AR core.
We also present an analysis of one of the flares it produced, SOL2020-04-09T09:32. 
The XSM analysis  indicates 
isothermal temperatures reaching 6 MK. The lack of very high-T emission is 
confirmed by AIA. We find excellent agreement
between the observed XSM spectrum and the one predicted using an AIA DEM analysis.
In contrast, the XRT Al-Poly / Be-thin filter ratio  
gives lower temperatures for the quiescent and flaring phases. 
We show that this is due to the sensitivity of this ratio to low temperatures,
as the  XRT filter ratios predicted with a DEM analysis based on
EIS and AIA gives values in good agreement with the observed ones.
\end{abstract}

  
\section{Introduction}
\label{sec:intro}

There is ample literature on observations and modelling of relatively large flares,
of GOES X-class C and above, but comparatively little on the 
weaker flares in  active region (AR) cores, despite the fact that they are  much more frequent.
We define here microflares as those events
of GOES class A or below, not to be confused with larger flares, often also called
microflares.
The physics of microflares remains elusive, as 
key spectroscopic observations to study the evolution of
5--10 MK plasma have been lacking, as discussed in the review by
\cite{delzanna_etal:2021_sxr}. The review points out
the need for high-resolution,
high-sensitivity  spectral imaging in the soft X-rays (100--150~\AA) to study microflares.

An earlier statistical study based on  GOES X-ray and Yohkoh
Bragg Crystal Spectrometer (BCS) observations   found a relationship between
  temperatures ($T_{\rm max}$) and X-ray emission measures (EM) at the peak of the X-ray emission, with lower-class  microflares generally having lower temperatures \citep[][]{feldman_etal:1996}. 
 A-class microflares had temperatures around 5 MK i.e. close to the `basal' 2--3 MK values of  active region cores \citep[cf][]{delzanna_mason:2018}. 
 One limitation of this and 
 similar studies  was the assumption that the plasma was isothermal.

 Yohkoh Soft X-ray Telescope (SXT) broad-band observations of
 microflares in AR cores were used to obtain isothermal temperatures ranging between 4--8~MK
 \citep{shimizu:1995}, lasting for 2--7~min.
 
The sensitivity of the SphinX X-ray irradiance spectrometer,  on board the CORONAS-PHOTON mission 
 was higher than that of the earlier GOES X-ray monitors,
and showed that microflares are indeed very common. Some studies of the general
characteristcs of microflares, produced from the Bremsstrahlung emission
and the isothermal assumption \citep[see, e.g.][]{kirichenko_bogahev:2017},
found a relation between $T_{\rm max}$ and EM for the smaller events
that was different to that found by \cite{feldman_etal:1996}. 

Studies of a few microflares,  based on NuSTAR observations of the
Bremsstrahlung emission and again the isothermal assumption  
\citep[see, e.g.][]{hannah_etal:2019,cooper_etal:2020ApJ...893L..40C,duncan_etal:2021} have been published.
Temperatures were in the range 4--8 MK and little evidence of non-thermal emission was generally found.

 Several questions arise from  previous studies of flares. 
First, is there a relation between the highest temperatures in the post-flare loops and the X-ray class ? 
The above-mentioned \cite{feldman_etal:1996} study clearly indicates that larger
flares have higher temperatures at their peak X-ray emission, but there are e.g.
B-class flares that reach very different maximum temperatures. 
In the example discussed below, a B-class flare did not reach 7 MK, while in the 
textbook case of another  B-class flare, observed by Hinode EIS \citep{culhane_etal:2007}, a peak temperature of 12 MK was reached \citep{delzanna_etal:2011_flare}.
Perhaps such differences are related to the physical size of the flaring volume.
A study of larger (GOES C8 class) flares by \cite{Bowen2013} indeed indicated that 
the peak temperature (in the 10--20 MK range) is inversely correlated with flare volume.

Second, is the isothermal assumption reasonable for small flares?
To address this question, detailed spectroscopic observations are needed.
Some information was obtained by \cite{mitra-kraev_delzanna:2019}
with the analysis of Hinode EIS observations of a microflare.
This event was composed of a bundle of strands being activated
at similar times, with sizes that appeared resolved at the AIA resolution,
1\arcsec. The EIS spectra indicated that these strands had relatively low peak temperatures (about 4--5 MK) and were nearly isothermal in their cross-section, 
although  the 5--10 MK  temperature range was not well constrained
\citep[see also][on the limitations in this range]{winebarger_etal:2012}.
The EIS results were consistent with the broad-band observations by the 
Solar Dynamics Observatory/Atmospheric Imaging Assembly (SDO/AIA)~\citep{lemen_etal:2012}
and Hinode XRT.
Multi-wavelength observations of several other microflares were studied by
\cite{mitra-kraev_delzanna:2019}, indicating that the example provided was a 'typical' case.

Another science questions about microflares relates to their
 possible contribution to the basal heating in AR cores.
 This basal emission has peak near-isothermal emission around 3 MK and does not
significantly change its characteristics over time 
\citep{delzanna:2013_multithermal, delzanna_etal:2015_emslope}. 
Detailed studies about the frequency of microflares have been lacking. They have typical lifetimes
of about 10 minutes as seen in the EUV \citep{mitra-kraev_delzanna:2019}.
The older X-ray observations of AR cores from
the Solar Maximum Mission  (SMM)  X-ray polychromator \citep{acton_etal:1980}
Bent Crystal Spectrometer (BCS) also showed common weak flaring
emission on those timescales  \citep{delzanna_mason:2014}, so it is possible that 
they were the same type of events. The  \ion{Mg}{xii} images from CORONAS-PHOTON 
also show a peak in the distribution of AR flaring events around 10 minutes
\citep{reva_etal:2018}, although it is not clear if these are the same
features (temperatures are uncertain  as
\ion{Mg}{xii} can be formed between 5 and 25 MK).

Larger flares such as the B-class textbook case \citep{delzanna_etal:2011_flare} show  quick
chromospheric evaporation and filling of the flare loops with hot (above 10 MK)
plasma, with a slow cooling phase where the plasma radiates at all
temperatures whilst draining back to the chromosphere.
In contrast, microflare loops are observed to cool down to nearly the background 3 MK, but
then drain/disappear very quickly and do not show emission at lower temperatures \citep{mitra-kraev_delzanna:2019}.
This suggests that microflares do not contribute to the basal
emission in AR cores. 

The chemical 
abundances of the flaring plasma also hold key information related to the heating and cooling.
The chemical abundances of low- FIP elements such as Fe, Si, relative to those
of high-FIP ones such as S,O,Ne 
were found to be increased by a factor of about 3.2
(with respect to the photospheric values) for 
AR cores, regardless of their size and age  \citep{delzanna:2013_multithermal,delzanna_mason:2014}.
This is the so-called FIP effect.

The FIP effect has gained a lot of interest in recent literature.
It is most likely caused by ion-neutral separation in the chromosphere, 
due to the Ponderomotive force of trapped Alfven waves in 
magnetically-closed loops  \citep[see references in][]{laming:2015}.
Observationally, there has been a wide range of different 
chemical abundances measured in coronal plasma, as reviewed by 
 \cite{laming:2015,delzanna_mason:2018}. 
 Regarding large flares,
there is ample evidence that the flare loops have near-photospheric 
abundances. We do not have measurements of abundances in microflares,
but if they were available and were also photospheric, that would also 
provide an indication that microflares do not contribute to the basal
emission in AR cores. 

The Solar X-ray Monitor (XSM) \citep{vadawale_etal:2014,shanmugam_etal:2020} 
onboard Chandrayaan-2 has been providing disk-integrated
spectra in the 1 -- 15 keV energy range since September 2019, providing
temperatures, emission measures, and absolute chemical abundances of a few elements,
with timescales of the order of a minute.
XSM has a much greater sensitivity than previous GOES X-ray monitors, and is able to
observe microflares down to A 0.01 class. 
A discussion of sub-A class flares during the minimum of solar cycle 24,
and which occurred outside of ARs was presented in \cite{xsm_microflares_2021}.
The instrument has also been used to study the X-ray emission of the global quiet Sun corona,
which was found to be dominated by X-ray bright points. A temperature close to
2 MK and a  coronal FIP bias of about two were found \citep{xsm_XBP_abundance_2021}.

XSM does not have enough sensitivity to measure chemical abundances 
for individual microflares, but is able to do so for B-class flares.
\cite{mondal_etal:2021} presented a time-resolved spectroscopic analysis of several
B-class flares occurring in AR cores. During the peak of the X-ray emission, temperatures
reached 6--8 MK, whilst the chemical abundances of Mg, Al, and Si  decreased
towards their photospheric values.  Similar decreases in the abundances were found at the peaks of larger flares by the Miniature X-Ray Solar Spectrometer (MinXSS) CubeSat \citep{Woods2017,Moore2018}. During the gradual decay phase, the XSM measurements found that
the chemical abundances  returned to their pre-flare coronal values.
  The MinXSS and XSM measurements are the first  observations of this kind,
  providing new observational constraints to models of the atmosphere which 
include element fractionation, such as those of \cite{laming:2017}.

To complement the XSM observations, in particular to understand the spatial
distribution of the X-ray emission observed by XSM,
we have run several multi-wavelength campaigns, including
Hinode HOP 396.  For this paper, we have selected 
observations during March/April 2020, when a single AR (NOAA 12759) 
crossed the solar disk and produced
many microflares and several B-class flares. 
The main aim of this paper is to present a multi-wavelength 
analysis of Hinode XRT and EIS observations, combined with
XSM and SDO/AIA during the AR disk passage, to establish the active region core 
temperatures and abundances, and discuss the temperatures of one of the 
flares, as measurable by these instruments.

\section{Observations and Data Analysis}\label{Observation}

We have analysed a large number of microflares observed by XSM during the first period
of operations and searched for suitable Hinode observations.  
We have selected for presentation here observations of
AR 12759  during its disk crossing, as a direct comparison
between the XSM irradiances and the signal measured by the other instruments
is straightforward, as this was the only AR on the disk. Also, we selected 
one of the flares to study, which occurred on Apr 9, as it had
XRT coverage. 
EIS observed the AR in several instances during
its quiescent phase, but did not observe any flares. 

\begin{figure*}[htbp!]
  \centering
\includegraphics[width=.6\linewidth]{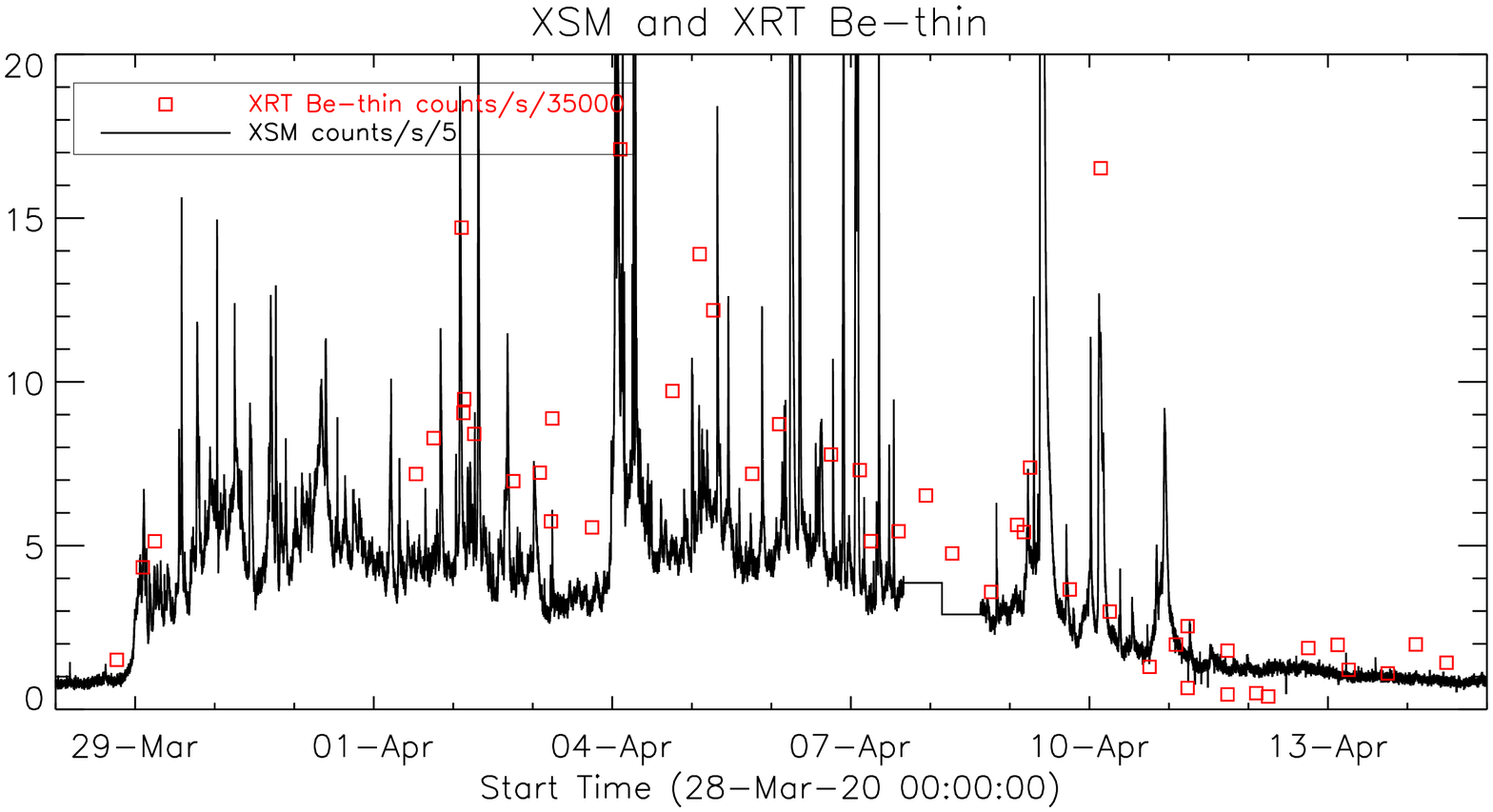}
\includegraphics[width=.6\linewidth]{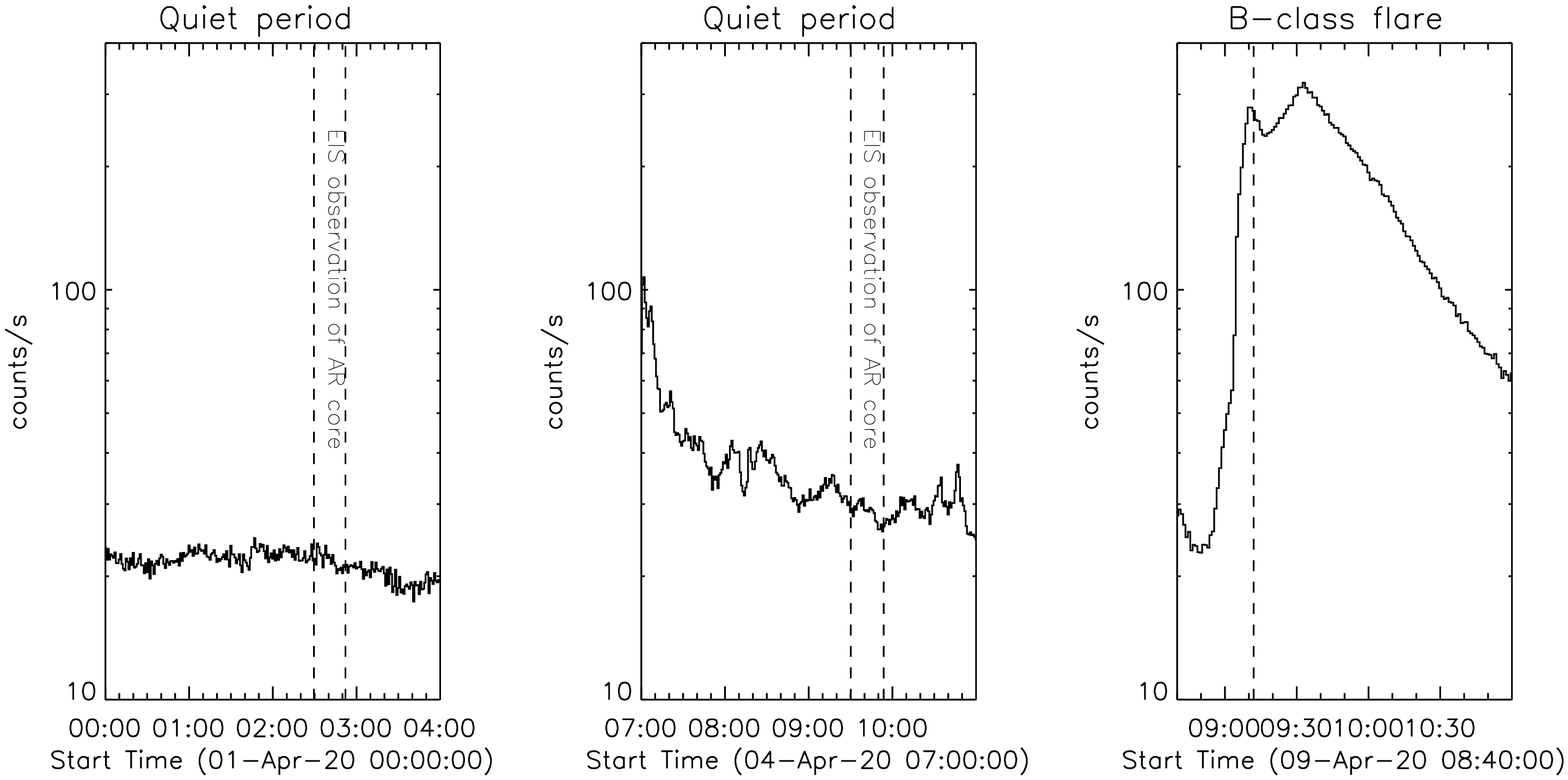}
\caption{Top: the XSM { 1--15 keV}  light curve (count rates averaged over 1 minute
  and scaled by a factor of 5) during the
    disk crossing of AR 12759. The red boxes indicate the scaled count rates from the full-disk Be-thin
    XRT filter.
    Bottom: the XSM light curves during three key periods selected for further analysis (see text). }
\label{fig:BClassFlare_lc}
\end{figure*}

\subsection{XSM}

{
XSM observes the Sun from a lunar orbit. It provides disk-integrated solar spectra at every second in the energy range of 1-15 keV with an unprecedented energy resolution of 180 eV at 5.9 keV~\citep{Mithun_2020SoPh..295..139M}. The
observed raw (level-1) spectra at every second are stored in a day-wise data file. Since the visibility of the Sun varies within the XSM field of view and the Sun also gets occulted by the moon, the level-1 data contains both solar spectra as well as non-solar background. We used XSM Data Analysis Software~\citep{xsmdas_2021} to generate the level-2 science data product for the Solar observations by selecting the Good Time Intervals (GTI) based on the observing geometry and a few other instrumental parameters. To do a spectral analysis of the XSM spectra we used the ``chisoth'' model ~\citep{mondal_etal:2021}. This is a local model of X-ray spectral fitting package (XSPEC)~\citep{Arnaud_1996}, meant for the spectral fitting of the observed X-ray spectrum.
The ``chisoth'' model calculates the synthetic photon spectrum from a spectral library, generated by using  CHIANTI v.10.  The model takes the logarithm of the temperature, abundances of the elements with  Z$=$2  to  Z$=$30,  and  the  volume  emission  measure as input parameters.}

Fig.~\ref{fig:BClassFlare_lc} shows the XSM 2--15 keV lightcurve
  during the    disk crossing of AR 12759.
  By March 28 the outer part of the AR appeared outside the
  east limb, when the X-ray 
 `background' increased. During its disk crossing, this AR 
  (the only one on the visible disc) produced  many microflares.
  A few of the larger, B-class flares have already been analysed 
  by \cite{mondal_etal:2021}. In all those cases, the 
  behaviour of the chemical abundances was similar, i.e. the abundances were closer
  to photospheric values during peak X-ray emission.
    By April 12 the core of the AR was behind the west limb.
  This paper focuses on
    an analysis of the quiescent emission from the AR during its disk 
    crossing on the Apr 1, 4, 8 and the B-class flare that peaked on Apr 9 around 9:12 UT.
    The  XSM lightcurves for  Apr 1, 4, 9 are shown in Fig.~\ref{fig:BClassFlare_lc}
    (bottom plots, on Apr 8 XSM did not observe).

\subsection{Hinode XRT}

We processed the XRT images using the standard SolarSoft procedures,
paying particular attention in removing saturated images and
taking into account the pointing information stored in the
XRT database (SolarSoft dbase).
To obtain the temperatures, we used CHIANTI v.10 \citep{chianti_v10} atomic data and calculated the responses.
The Al poly / Be thin filter ratio is essentially insensitive to different chemical abundances \footnote{See http://solar.physics.montana.edu/takeda/xrt\_response/xrt\_resp\_ch1000.html}.

We have analysed full-Sun XRT synoptic observations during the
AR disk crossing. The averaged count rates in the Be-thin filter
for a selection of dates are shown in Fig.~\ref{fig:BClassFlare_lc}
as boxes, scaled by a normalization factor.
There is generally good agreement in the variability of the
XRT signal in this filter and the XSM count rates, as expected.
This is because the XSM sensitivity is close to that of this XRT
filter. The XRT Be-thin full-disk images are also very useful for
checking that  AR 12759, even during quiescence, is
dominating the full-disk signal. This means that the XSM signal is
also dominated by the active region, making a direct comparison
between XSM and the other instruments (XRT, EIS, AIA) possible.

There are also many XRT partial-frame observations of the
active region during its disk crossing, with the same filters.
We have used one set for a direct EIS/XRT cross-calibration. 

For each XSM flare we searched for available XRT observations.
As microflares last about 10 minutes, it was essential to
obtain an XRT  cadence of about a minute, which was obtained with a
reduced FOV and the use of two filters.
In many instances, the automatic exposure control (AEC), a safety
measure for the instrument, was too slow to reduce the exposures,
and the  XRT images during the peak of the microflares were saturated
(the standard AEC works relatively well for larger flares, as they
have a longer duration). 
One observation which we could use for analysis was the
flare on Apr 9, produced by AR 12759, where
XRT used the Al-poly and Be-thin filters.

\begin{figure*}[htbp!]
  \centerline{
  \includegraphics[width=0.25\linewidth]{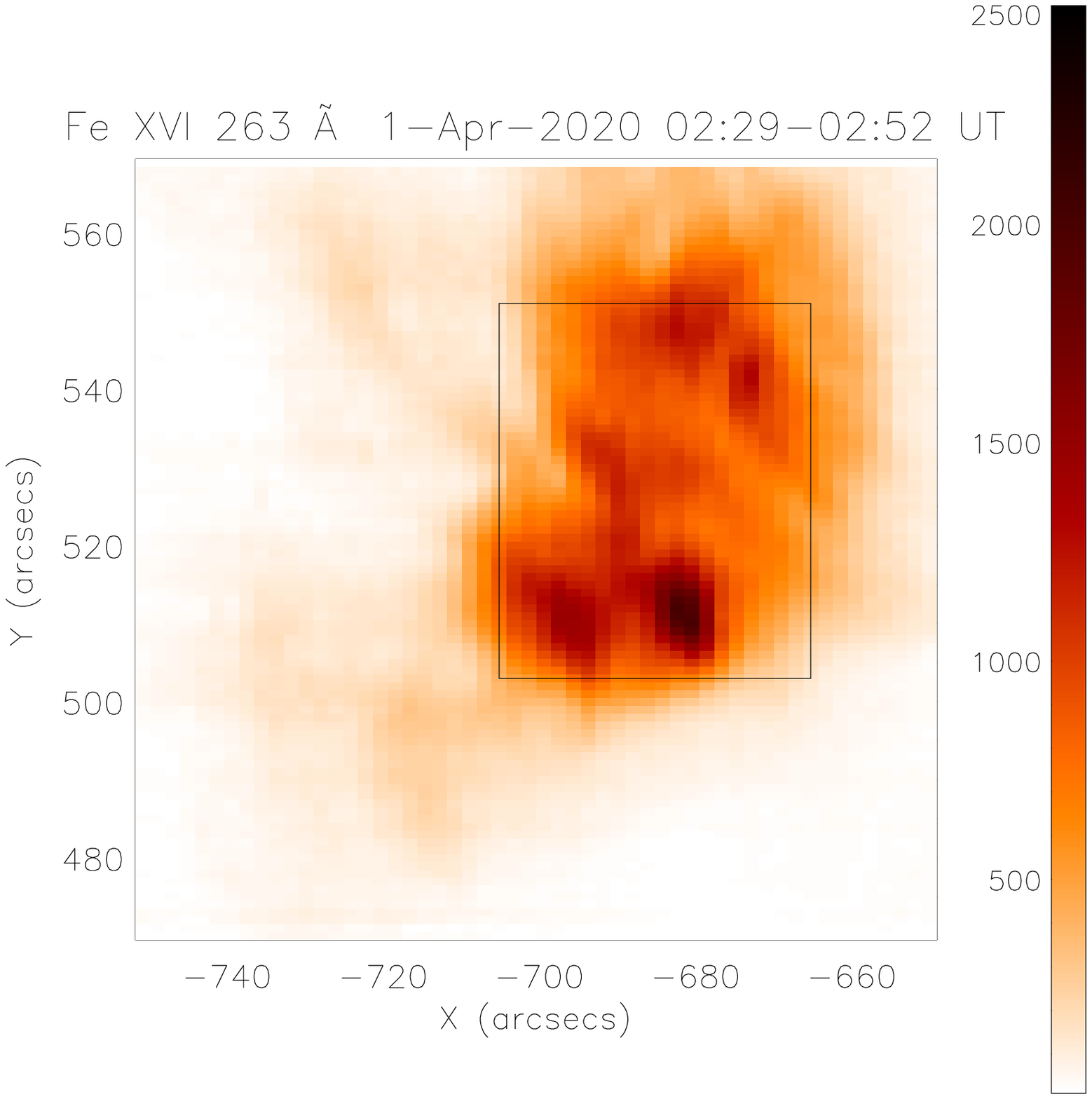}
   \includegraphics[width=0.25\linewidth]{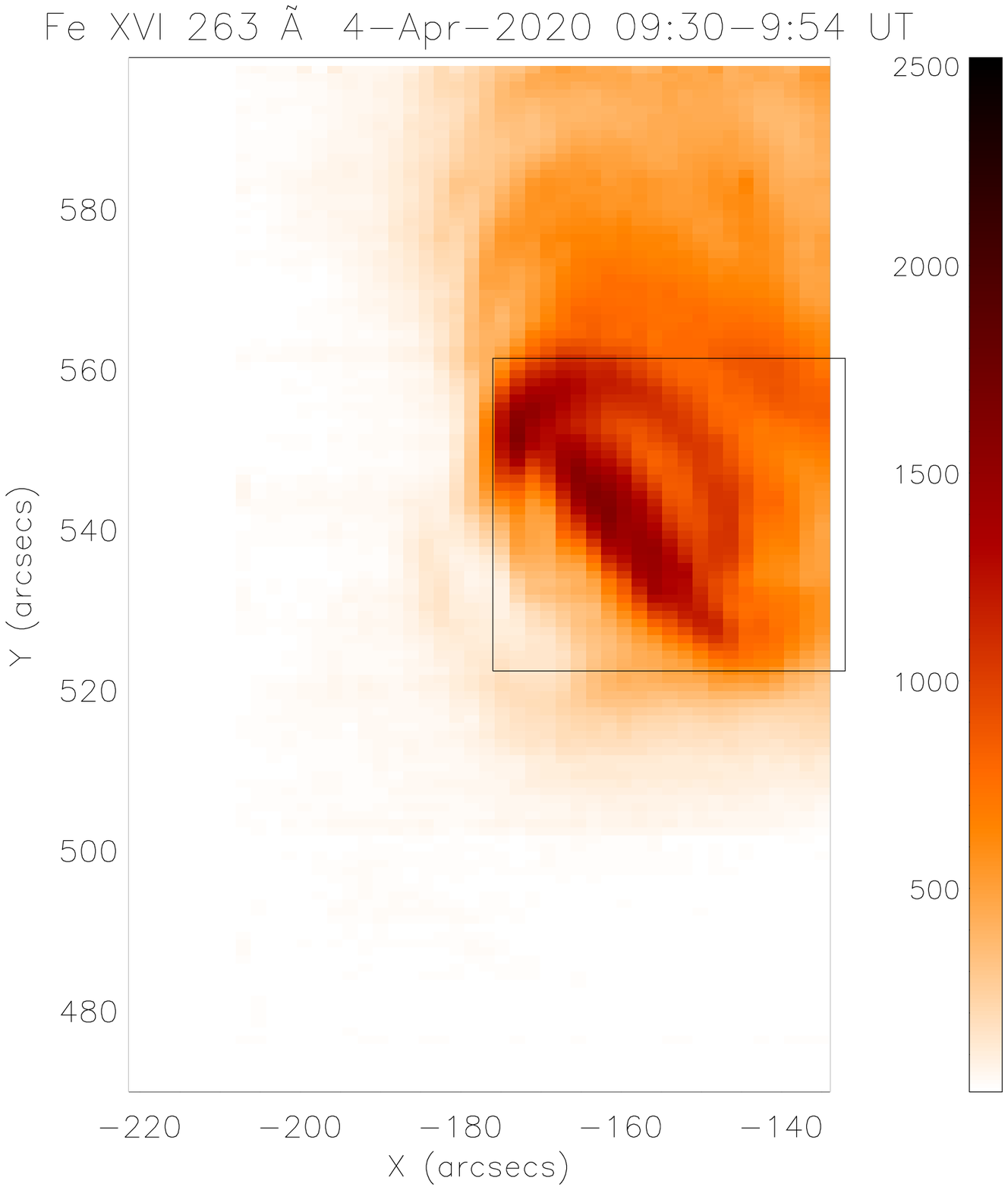}
    \includegraphics[width=0.25\linewidth]{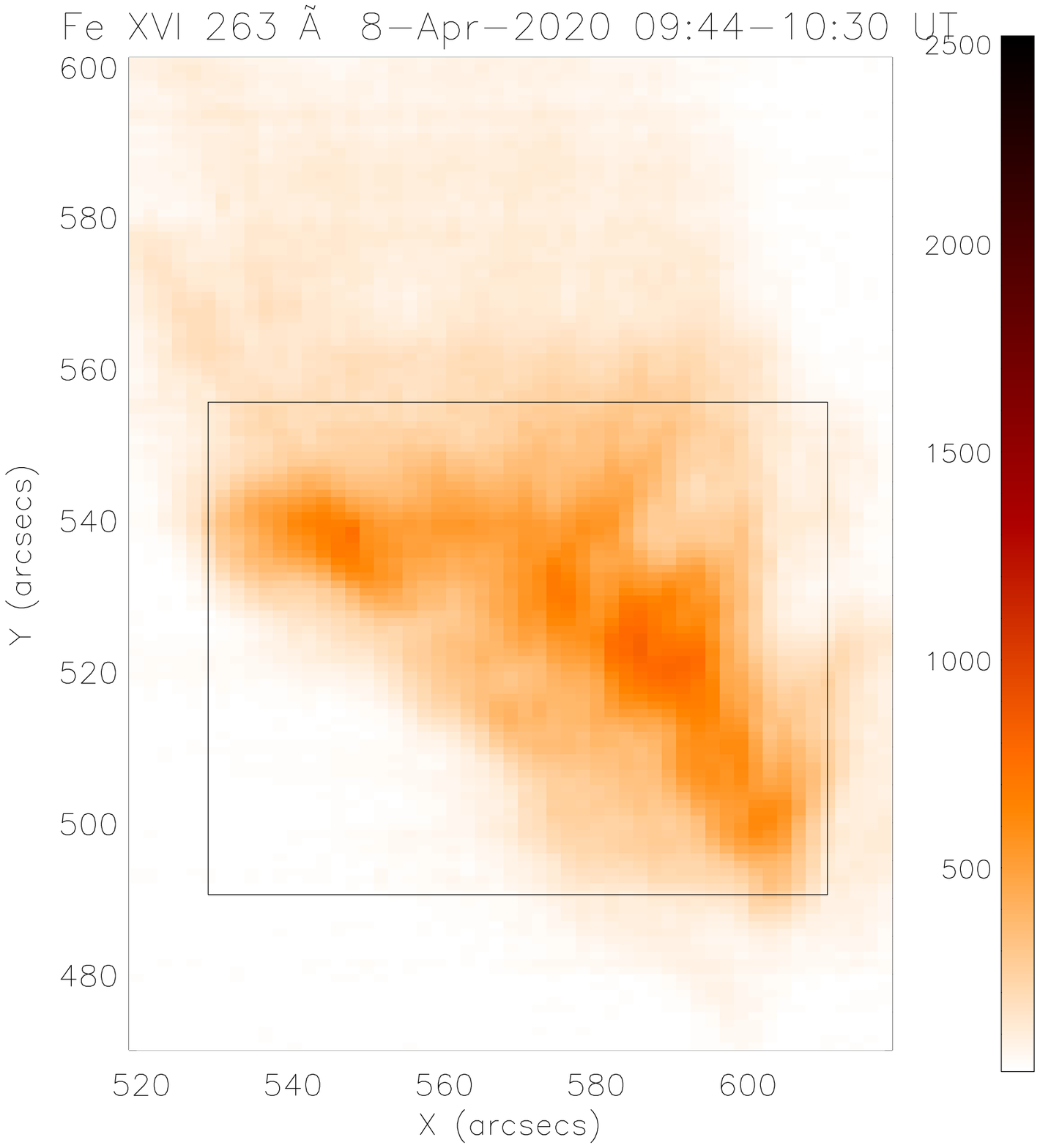}}
  \centerline{
    \includegraphics[width=0.25\linewidth, angle=0]{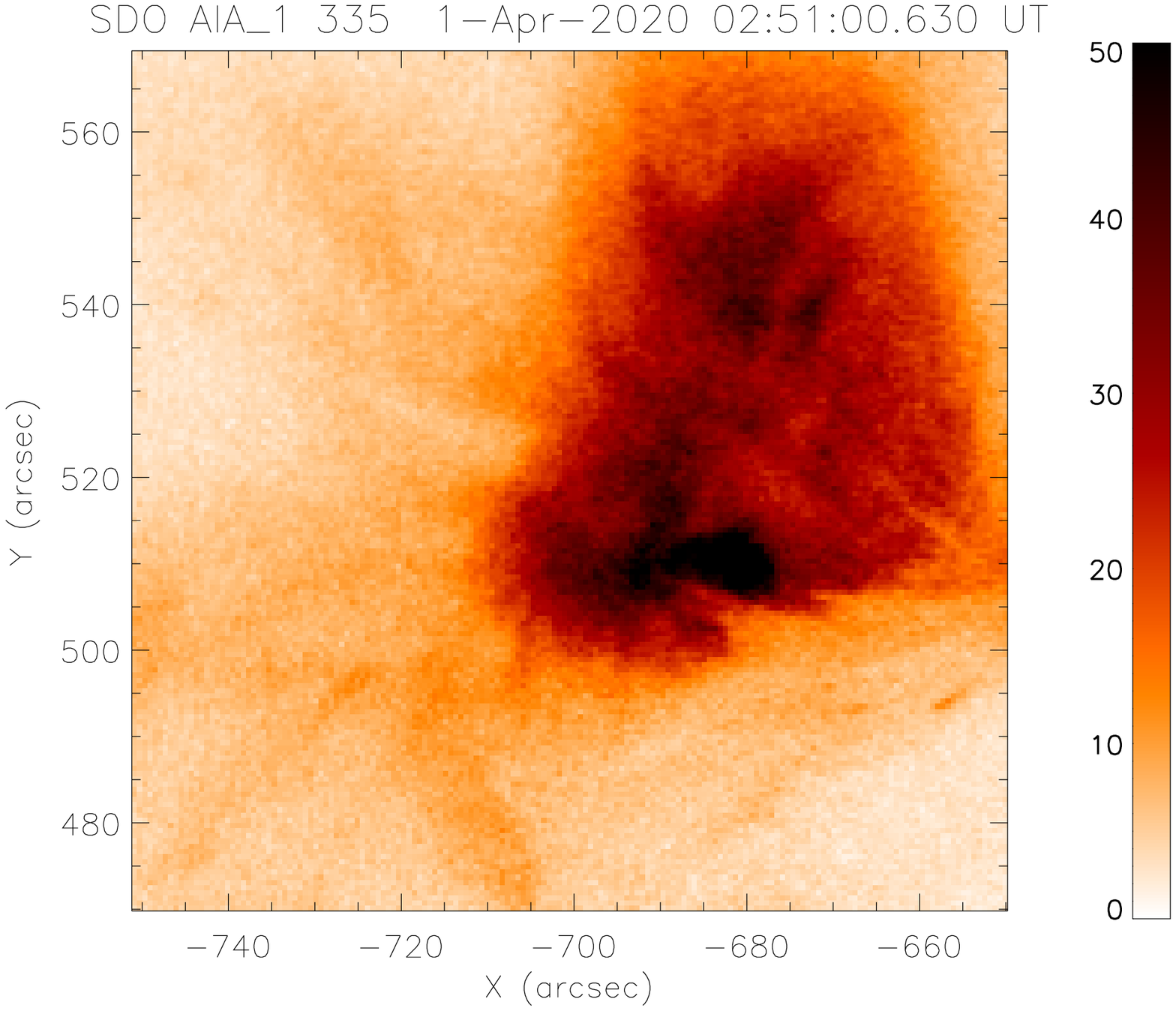}
    \includegraphics[width=0.245\linewidth, angle=0]{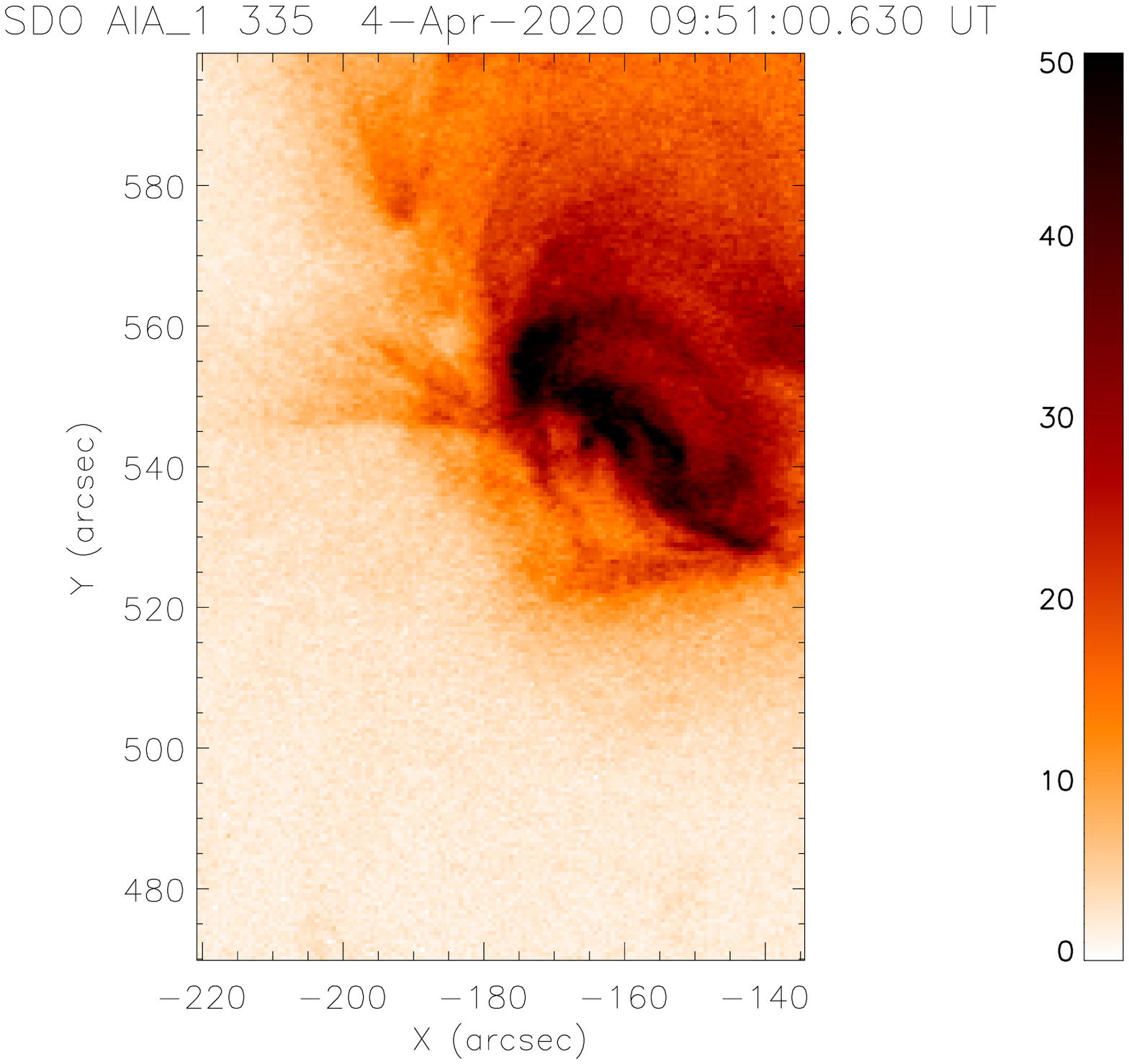}
    \includegraphics[width=0.25\linewidth, angle=0]{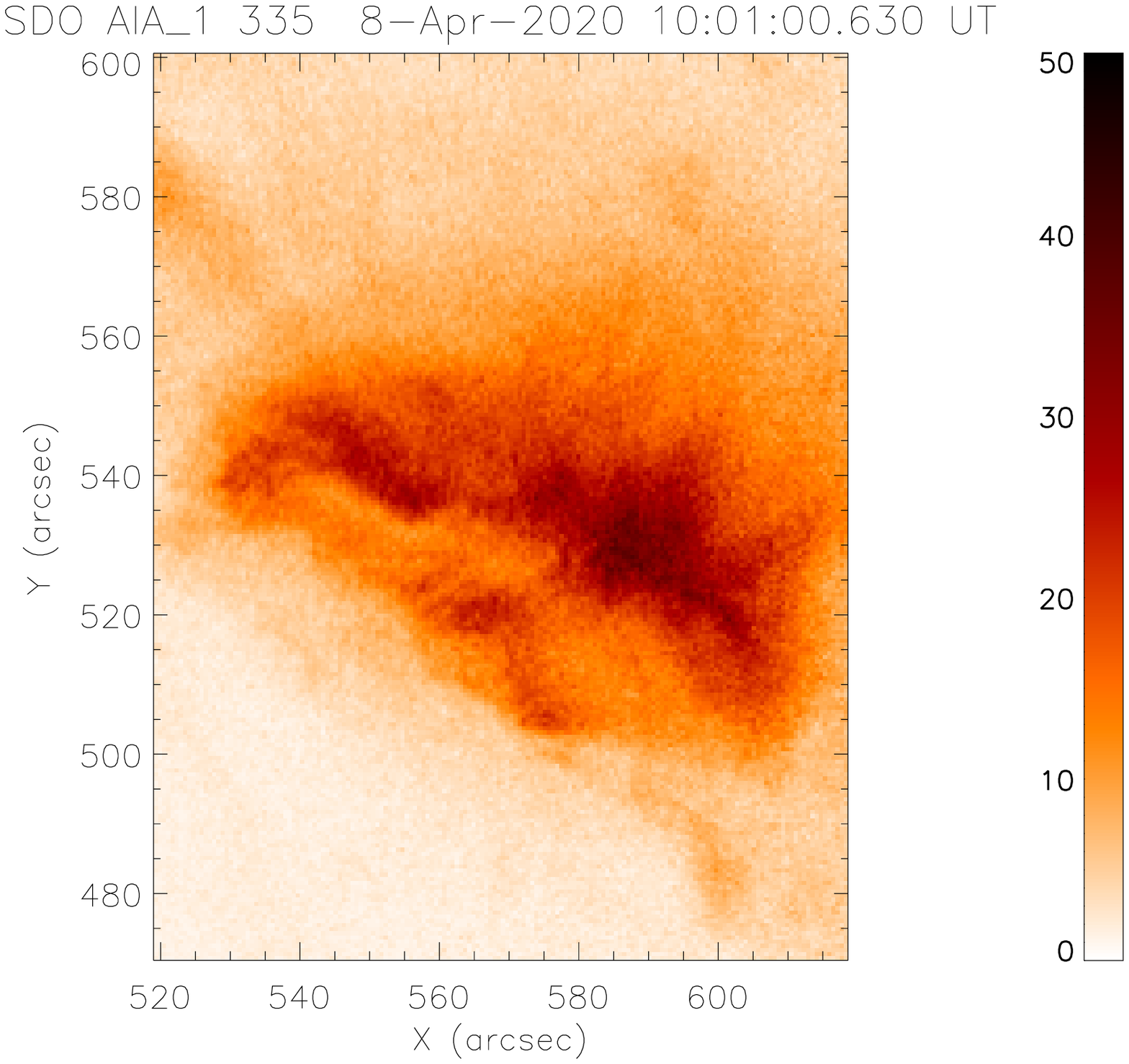}
    }
     \centerline{
         \includegraphics[width=0.25\linewidth, angle=0]{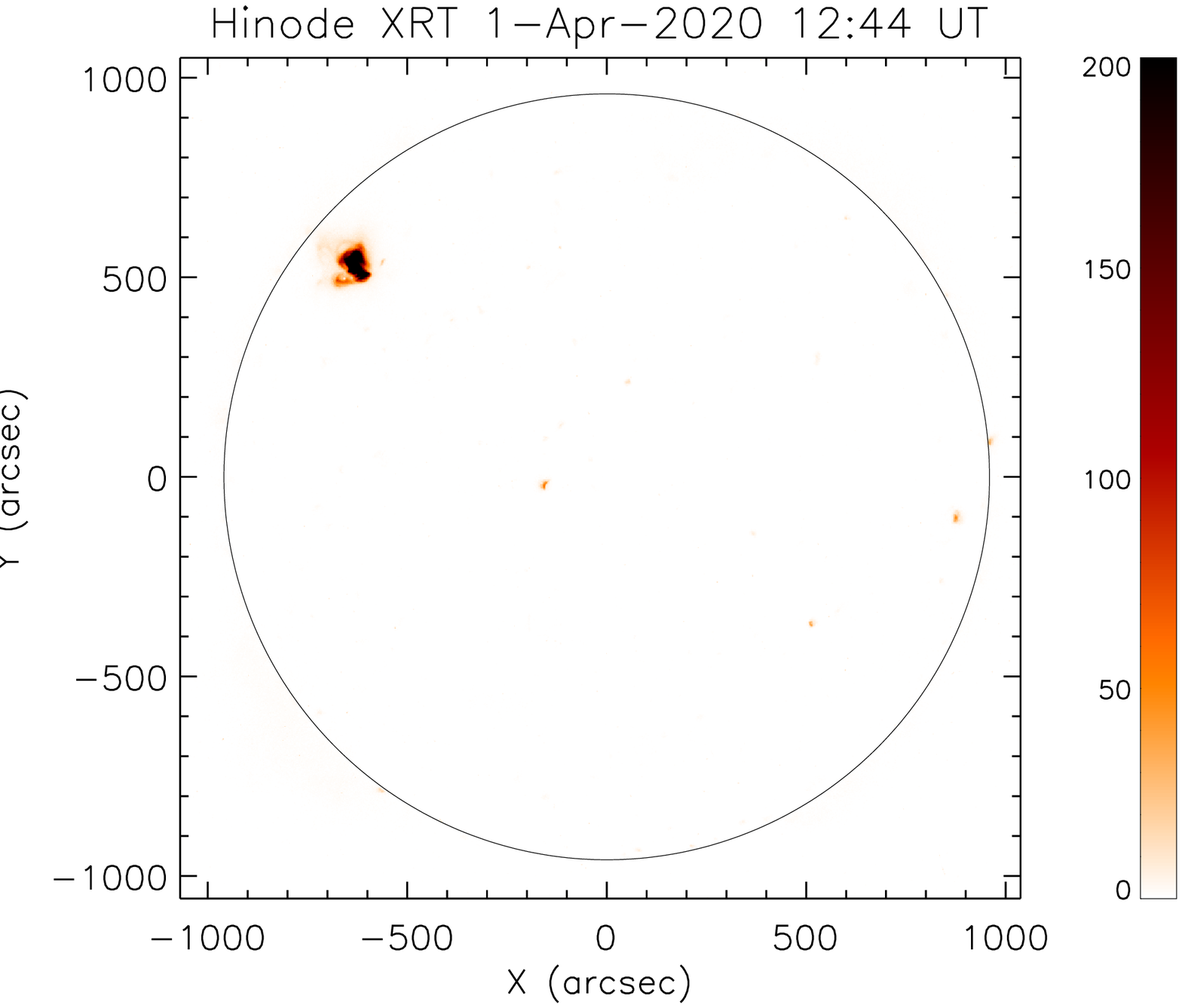}
          \includegraphics[width=0.25\linewidth, angle=0]{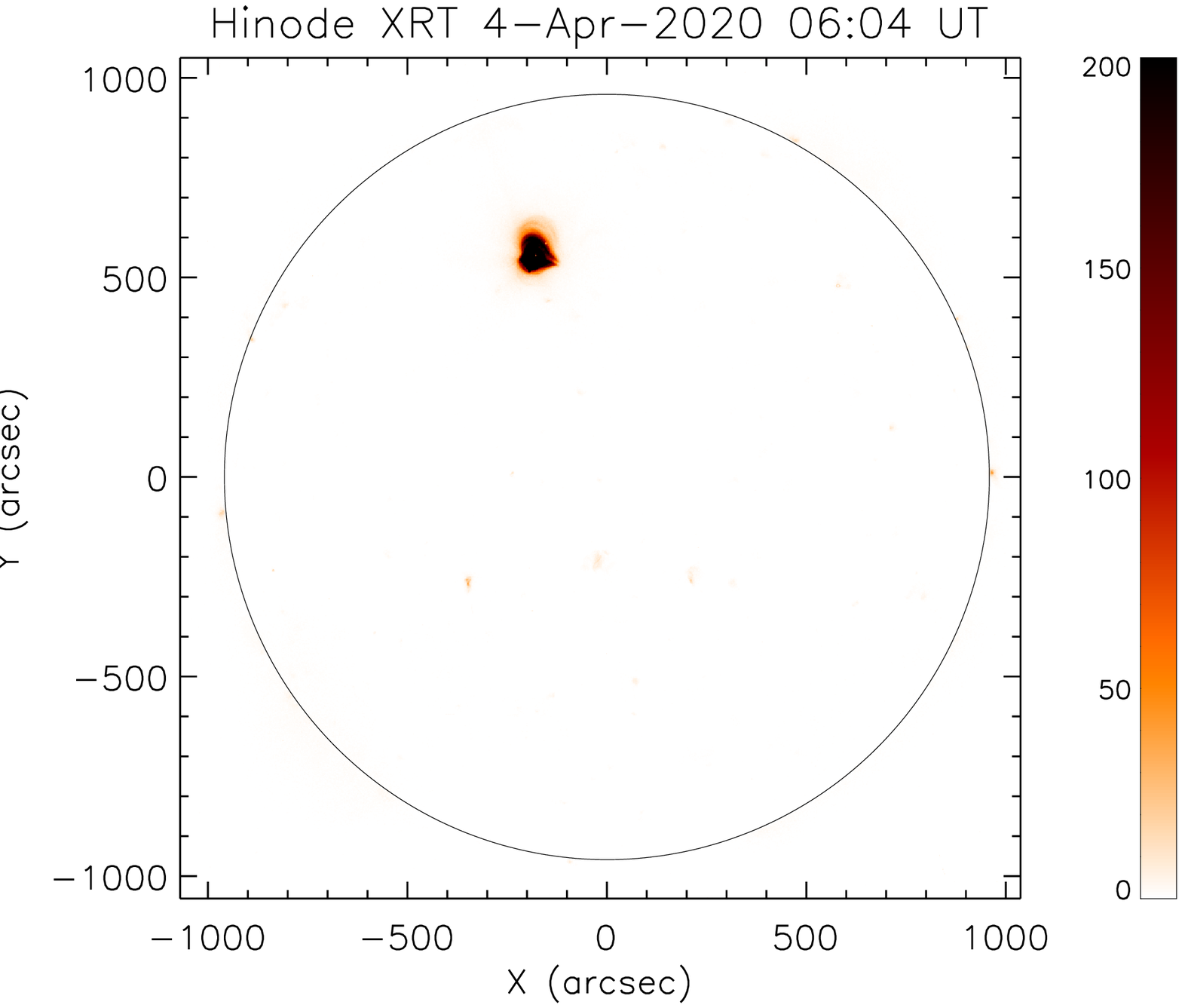}
           \includegraphics[width=0.25\linewidth, angle=0]{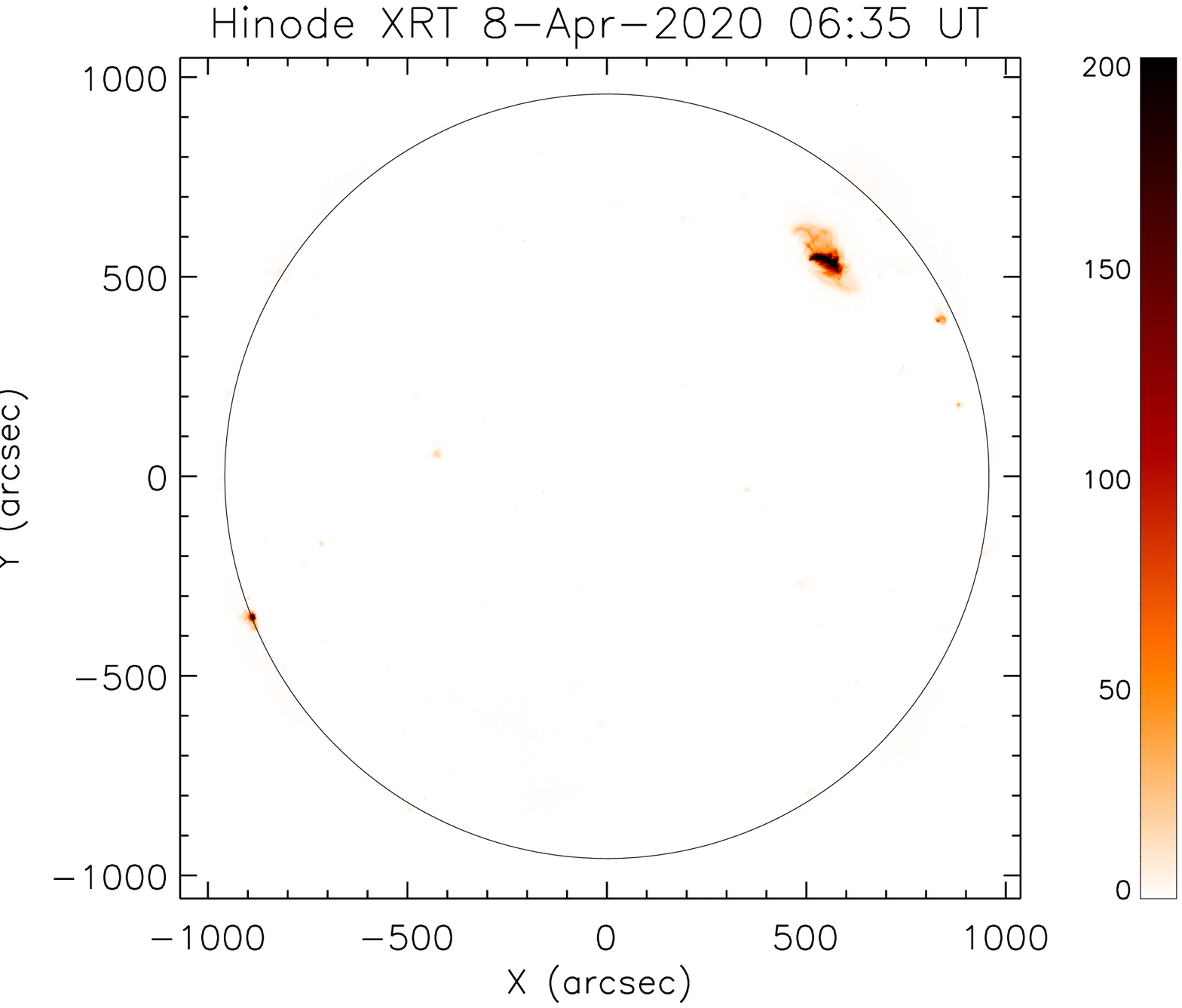}
     }
    \caption{Top row: selected regions (boxes) of the AR 12759 core EIS 
    observations  during its disk crossing on 2020 Apr 1, 4, and 8.
  The images show the integrated radiances in the Hinode EIS
  Fe XVI 263~\AA\ line, in calibrated units.
Middle row: SDO AIA 335~\AA\ images during  the EIS observations, in DN/s.
Bottom row: XRT Be-thin images (DN/s) at times close to the EIS observations.
All images are  negative.
}
\label{eis}
\end{figure*}

\subsection{Hinode EIS}

For each XSM flare we searched for available EIS observations,but
found none. However, there is an excellent set of observations of 
AR 12759 during its disk crossing, allowing us to check the
active region evolution. 

To process EIS data, we used custom-written software which
mostly follows the standard SolarSoft programs, with the exception that
the bias is  subtracted while doing the line fitting, and the particle hits are
removed by averaging procedures along the slit direction. 

The  main issue we faced in the analysis is related to the EIS radiometric calibration,
and its variations in time and wavelength.
\cite{delzanna:13_eis_calib} presented a significant revision of the ground
calibration, with an additional time-dependent decrease in the sensitivty
of the long-wavelength channel, however this was only  valid until 2012, and indicated
that  further wavelength-dependent corrections with time would eventually be needed.
A long-term programme was therefore started  to analyse data
after 2013 and provide a new radiometric calibration.
The preliminary  results of this study (Del Zanna and Warren, 2022, in prep) are used here. More details 
are provided in the Appendix.

\subsection{SDO/AIA}

As EIS did not observe the flare on April 9, we used SDO AIA data
to study the thermal structure of the flare.
We have used Mark Cheung’s Differential Emission Measure (DEM) code
\citep{Cheung2015}.
As input we used averaged images over 36 s (images at 3 consecutive times)
from the six EUV AIA channels (94, 131, 171, 193, 211, 335~\AA) to derive EM maps.
We took a minimum log $T$ [K] = 5.7 and 13 temperature bins of width 0.1 dex.
The SDO/AIA response functions  were calculated using CHIANTI version 10 considering
a constant pressure of 10$^{15}$ cm$^{-3}$ K and
chosen abundances.
We used the effective areas with the estimated degradation
in the various channels as available within SolarSoft, noting
that such degradation was extrapolated from the last
calibration sounding rocket flight which flew in 2018.
Hence, some additional uncertainty is present.

\section{Results}

\begin{figure}[ht!]
  \centering
  \includegraphics[width=.8\linewidth, angle=-90]{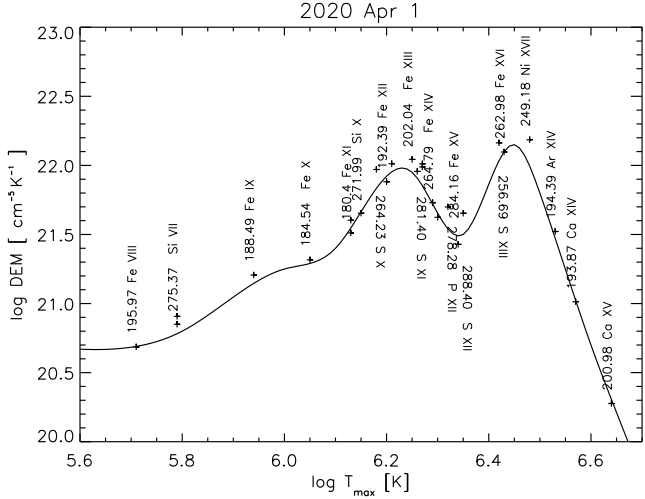}
  \caption{DEM for the AR core observed on 2020 Apr 1, obtained from Hinode EIS.
  The points are plotted at the temperature $T_{\rm max}$, 
 and at the theoretical vs. the observed
intensity ratio multiplied by the DEM value.}
\label{fig:eis_dem}
\end{figure}

\def\baselinestretch{1.}
\begin{table}[!htbp]
  \caption{List of  emission lines from the 2020 Apr 1 active region core.
  $\lambda_{\rm obs}$ (\AA) is the measured wavelength.
DN is the number of total counts in the line, while $I_{\rm obs}$
is the calibrated radiance (phot cm$^{-2}$ s$^{-1}$ 
arc-second$^{-2}$) obtained with our new EIS calibration.
$T_{\rm max}$ and $T_{\rm eff}$ are the maximum and effective temperature (log values in K),
$R$ the ratio between the predicted and observed radiances, 
Ion and $\lambda_{\rm exp}$ (\AA) the main contributing line, and 
$r$ the fractional contribution to the blend.}
\begin{center} 
\scriptsize
\begin{tabular}{@{}lllllllll@{}}
 \hline\hline \noalign{\smallskip}
  $\lambda_{\rm obs}$  & DN & $I_{\rm obs}$   &  $T_{\rm max}$ & $T_{\rm eff}$  & $R$ & Ion & $\lambda_{\rm exp}$   &  $r$ \\
 \hline \noalign{\smallskip}

 195.98 &      904 & 13.9 &  5.71 &  5.90 &  1.01 &  \ion{Fe}{viii} &  195.972 & 0.85 \\ 
 
 275.36 &      361 & 52.0 &  5.79 &  5.92 &  0.85 &  \ion{Si}{vii} &  275.361 & 0.99 \\ 
 
 280.73 &       62 & 17.4 &  5.79 &  5.93 &  0.75 &  \ion{Mg}{vii} &  280.742 & 0.91 \\

 188.49 &     1920 & 61.2 &  5.94 &  6.05 &  0.87 &  \ion{Fe}{ix} &  188.493 & 0.90 \\ 

  197.86 &     1580 & 29.1 &  5.96 &  6.07 &  0.86 &  \ion{Fe}{ix} &  197.854 & 0.88 \\ 

 184.54 &     2610 & 194.0 &  6.05 &  6.14 &  0.94 &  \ion{Fe}{x} &  184.537 & 0.94 \\ 
 
 192.81 &     6300 & 122.0 &  6.13 &  6.18 &  0.88 &  \ion{Fe}{xi} &  192.813 & 0.94 \\ 
    
 180.40 &     2330 & 856.0 &  6.13 &  6.19 &  1.09 &  \ion{Fe}{xi} &  180.401 & 0.97 \\ 
  
 271.98 &     1040 & 119.0 &  6.15 &  6.21 &  1.02 &  \ion{Si}{x} &  271.992 & 0.97 \\ 
 
  192.40 &    13700 & 283.0 &  6.20 &  6.22 &  1.10 &  \ion{Fe}{xii} &  192.394 & 0.96 \\

 
 264.23 &      830 & 97.2 &  6.18 &  6.24 &  0.73 &  \ion{S}{x} &  264.231 & 0.98 \\

 202.05 &    10300 & 579.0 &  6.25 &  6.27 &  0.81 &  \ion{Fe}{xiii} &  202.044 & 0.96 \\ 
 251.95 &     1500 & 379.0 &  6.25 &  6.27 &  1.00 &  \ion{Fe}{xiii} &  251.952 & 0.97 \\ 
  
 188.67 &     1200 & 36.7 &  6.21 &  6.27 &  0.87 &  \ion{S}{xi} &  188.675 & 0.58 \\ 
                            &     &   &  &  &  &  \ion{Fe}{xii} &  188.679 & 0.11 \\ 
                            &     &   &  &  &  &  \ion{Fe}{xi} &  188.630 & 0.13 \\

 281.41 &      102 & 30.4 &  6.26 &  6.30 &  0.90 &  \ion{S}{xi} &  281.402 & 0.84 \\ 
                                 & &   &  &  &  &  \ion{Fe}{xi} &  281.367 & 0.13 \\

 285.83 &       83 & 46.0 &  6.27 &  6.31 &  0.73 &  \ion{S}{xi} &  285.823 & 0.97 \\ 
 
 
 
 246.89 &       98 & 39.6 &  6.27 &  6.31 &  0.70 &  \ion{S}{xi} &  246.895 & 0.98 \\

 264.78 &     6230 & 726.0 &  6.29 &  6.33 &  0.98 &  \ion{Fe}{xiv} &  264.789 & 0.93 \\ 
 

   
 211.32 &     2660 & 913.0 &  6.30 &  6.33 &  1.07 &  \ion{Fe}{xiv} &  211.317 & 0.97 \\ 
 
 278.27 &       31 & 7.1 &  6.32 &  6.36 &  0.69 &  \ion{P}{xii} &  278.286 & 0.95 \\ 
 
 288.40 &      131 & 111.0 &  6.35 &  6.39 &  0.70 &  \ion{S}{xii} &  288.434 & 0.98 \\

 284.15 &    10700 & 4620.0 &  6.34 &  6.40 &  1.15 &  \ion{Fe}{xv} &  284.163 & 0.98 \\ 
  
 194.42 &      432 & 7.3 &  6.53 &  6.42 &  0.89 &  \ion{Ar}{xiv} &  194.401 & 0.75 \\ 
                              &   &   &  &  &  &  \ion{Fe}{xi} &  194.442 & 0.13 \\

 256.68 &     2440 & 398.0 &  6.42 &  6.44 &  0.74 &  \ion{S}{xiii} &  256.685 & 0.98 \\ 
 
  262.98 &     2200 & 263.0 &  6.43 &  6.44 &  1.00 &  \ion{Fe}{xvi} &  262.976 & 0.96 \\

 249.18 &      680 & 227.0 &  6.48 &  6.47 &  0.65 &  \ion{Ni}{xvii} &  249.186 & 0.96 \\ 
 
 193.88 &     1110 & 19.7 &  6.57 &  6.48 &  1.02 &  \ion{Ca}{xiv} &  193.866 & 0.96 \\ 
 
 201.00 &      200 & 7.3 &  6.64 &  6.48 &  1.07 &  \ion{Ca}{xv} &  200.972 & 0.84 \\ 
 
 

\noalign{\smallskip}\hline                                   
\end{tabular}
\normalsize
\end{center}
\label{tab:list}
\end{table}

\subsection{EIS results for the quiescent AR core}

Hinode EIS observed AR 12759 during its disk crossing with several studies.
We analysed the ATLAS\_60 spectra, obtained with 60 s exposures on 
Apr 1,4, and 8 to obtain the DEM  and the FIP effect in the AR core.
There was no significant variability in the AR core during these three 
observations, as seen in the
XSM light curves (cf Fig.~\ref{fig:BClassFlare_lc}, bottom plots) 
and as observed from XRT.
For each observation, we selected an AR core region using the intensity
of the Fe XVI 263~\AA\ line, as shown in Fig.~\ref{eis}. 
The figure also shows AIA 335~\AA\ images, which are dominated in the AR core by 
Fe XVI, and full-Sun XRT Be-thin images, which show that the AR was dominating the 
signal from the Sun. The AIA full-disk images were used to coalign EIS.
 The AR on Apr 1 was close to the east limb,
  on the 4th it was close to meridian and on the 8th it was 
  close to the west limb.

We adopted the \cite{delzanna:2013_multithermal} chemical abundances,
where the FIP bias is 3.2 (low FIP elements increased compared to their
photospheric abundances) and used CHIANTI v.10 to obtain  DEMs. 
To calculate the line emissivities we used
a constant electron density of 2$\times$10$^9$  cm$^{-3}$, a typical 
average AR value which fits the main EIS density-sensitive line ratios.
The DEM was obtained  using the method
described in \cite{delzanna_thesis99}, where the 
$DEM$ is assumed to be a spline function, and considering only low-FIP ions.

The DEM for Apr 1st is shown in Fig.~\ref{fig:eis_dem}.
Those for Apr 4, 8 are similar and are shown in the Appendix.
The plasma distribution in temperature has two peaks, one around
1.5 MK, typical of the diffuse emission in the AR core, and
one around 3 MK, typical of the nearly isothermal hot AR core loops.
Both temperature structures have an FIP bias of about 3,
as the main lines are well reproduced within 20--30\% (the uncertainty in
the radiometric calibration).

Table~\ref{tab:list}
lists a small selection of the lines, showing a relatively good agreement
between the low FIP elements (e.g. Fe, Si, Ca) and the high-FIP element
argon.  There are several argon lines in the EIS spectra, but they are all 
extremely weak in AR core spectra 
and blended to some degree, some with unknown transitions.
The strongest line is an Ar XIV at 194.4~\AA, very close to 
a stronger unidentified line at 194.35~\AA.

Note that sulphur has a FIP of 10 eV, but can be used as
a proxy for the high-FIP elements.  In fact, in remote-sensing observations
its variations are in line with those of the high-FIP
elements. The sulphur EIS lines are strong and cover a wide temperature range, as
are the iron lines, so the Fe/S results are more accurate than Ar/Fe.
A few weak lines from other elements (e.g. K, P, Al) are
also present and confirm the chosen abundances. 

The Table also gives the total counts 
(DN) in the lines, as well as the calibrated intensities. 
It also indicates $T_{\rm max}$, the temperature where
the line contribution function $G(T)$ has a maximum,
and the effective temperature $T_{\rm eff}$:
\beq
T_{\rm eff} = \int G{\left({T}\right)}~
DEM{\left({T}\right)} ~T~dT / 
{\int G{\left({T}\right)}~DEM{\left({T}\right)}~dT} \quad ,
\eeq
that is  an average temperature more indicative of 
where a line is formed.

The DEM results of the AR core regions selected for Apr 4th and Apr 8th
are similar,  as shown in the Appendix,
although the AR core on the 8th had a lower emission in the
high temperature peak. We also find,  importantly,  that the 
chemical abundances of the quiescent AR core do not vary, regardless of the age and 
the fact that many flares occurred. This result agrees with our previous studies 
of several active regions.

\subsection{XSM results for the quiescent AR core}

\begin{figure}[htbp!]
  \centering
  \includegraphics[scale=0.5]{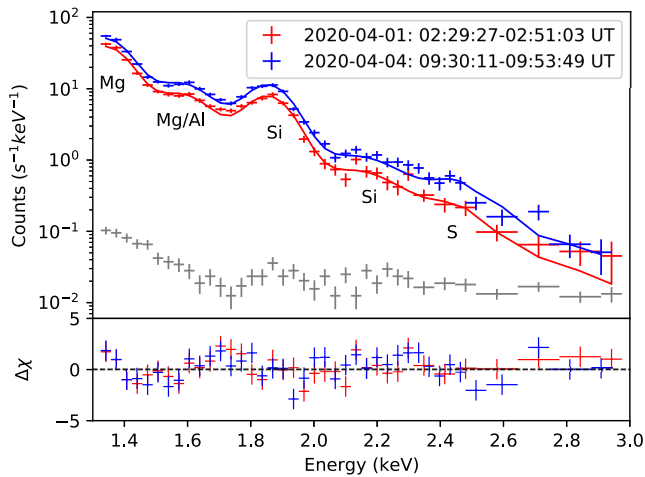}
  \caption{XSM spectrum during the quiescent phase of AR 12759 {(red and blue  points with error-bars)},
  together with the 
  best fit model spectrum {(red and blue solid lines)} and the background signal
  (grey). The lower panel shows the residuals between 
  observations and best fit model.}
\label{fig:XSM_ARspec}
\end{figure}  

\begin{deluxetable}{c c c c c c}
\tablecaption{AR parameters obtained from the XSM spectrum. The abundances of Mg,Al, Si are in dex. The EIS abundances are the reference values of \cite{delzanna:2013_multithermal} adopted here. The Feldman are those
listed by  \cite{feldman:1992}, while the photospheric ones are from 
\cite{asplund_etal:2009}. }
\label{ARfitpar}
\tablehead{
Date&log$_{10}$(T) & EM & Mg & Al & Si \\
&(K) & (10$^{46}$ cm$^{-3}$) &  &  &   
}
\startdata
01-04-2020&$ 6.47^{+0.01}_{-0.01} $&  $ 8.67^{+1.9}_{-1.9}$&  $       7.78^{+0.07}_{-0.06}$&  $  7.05^{+0.11}_{-0.12}$&  $  7.89^{+0.05}_{-0.04}$\\
04-04-2020&$ 6.50^{+0.02}_{-0.01} $&  $ 10.79^{+2.5}_{-1.7}$&  $       7.72^{+0.1}_{-0.08}$&  $  7.03^{+0.16}_{-0.17}$&  $  7.82^{+0.06}_{-0.06}$\\
 EIS  &  &  & (8.10) & 6.95 & 8.00 \\ 
 Feldman & & & 8.15 & 7.04 & 8.10 \\
 Phot. & & & 7.6 & 6.45 & 7.51 \\
\enddata
\end{deluxetable}

To measure the absolute FIP effect in the AR core, we have analyzed the XSM spectra of the quiescent AR on  April 1 from  02:29:27 UT to 02:51:03 UT, {and April 4 from 09:30:11 UT to 09:53:49 UT}, the timings of the EIS core observations. Due to the presence of the single AR on the solar disk, the XSM X-ray emission during this time was dominated by the AR core emission. We have fitted the XSM spectra with a single temperature model by considering the temperature, EM, and the abundances of Mg, Al and Si as a free parameters.
Figure~\ref{fig:XSM_ARspec} shows the XSM best fit model spectra
{(red and blue solid lined)} along with the observed ones {(red and blue points)}.
The best fitted parameters along with the 1-$\sigma$ error-bars are given in  Table~\ref{ARfitpar}.
The isothermal temperature of 3 MK agrees with the peak in the DEM
as measured by Hinode EIS. 
The absolute abundances of Mg, Al and Si are given in dex.
The silicon feature is the most prominent in the XSM spectra. The Si abundance is 
within 25\% the \cite{delzanna:2013_multithermal} AR core values. 
The Al abundance is also close, while the Mg abundance as measured by 
XSM is lower. 
We note that the relative Al and Si abundances were measured by 
\cite{delzanna:2013_multithermal} using lines formed at 
lower temperatures, between 1 and 1.5 MK, while the 
Mg abundance was estimated by assuming the same increase
over the photospheric value as Si (the Mg lines in the EIS spectra are formed
at much lower temperatures).
\cite{delzanna:2013_multithermal} estimated the absolute abundances assuming
a unitary filling factor and measuring the path length of an AR core loop, so the 
agreement with the XSM result is remarkable.

\subsection{XRT temperatures and EIS / XRT cross-calibration}

We have analysed several XRT observations of the AR (at non-flaring times), and
found consistently that the Be-thin/Al-Poly filter ratio
has a value around 0.13 in the AR core. This value is equivalent to a
temperature of about 2 MK, significantly lower than the 3 MK measured by XSM. 
Also, as shown below, the isothermal temperatures obtained from this XRT filter
ratio are significantly lower than those measured by XSM during the B-class flare.

We initially thought that this discrepancy could be due to an XRT calibration issue. 
There have been several reports in the literature about a discrepancy
between the estimated and measured XRT count rates
\citep[see, e.g.][and references therein]{mitra-kraev_delzanna:2019},
but not on possible problems in the relative calibration of the two channels.

We also know that a leakage of visible light in some of the filters
can significantly affect some filter ratios (but not the Be-thin).
Currently, it is possible to correct for this leakage by subtracting
specially selected calibration images from full-frames, but not
for partial frames. Also, the leakage can be more pronounced in partial
frames.
So, we first checked  Al-Poly full-disk calibration data, which is essentially a
flat field, and found that the variations between the center and
the edges of the FOV are within a few DN/s, hence are negligible.
Then we analysed
several full-frame XRT synoptic datasets, and obtained isothermal temperatures
for the AR core. The central part of the AR core is generally saturated in the
synoptic observations (either in the Al-Poly or the Be-thin), except the 1s exposures,
where the signal is often low, so combined images with different exposures need to be used. The full-disk filter ratios consistently indicate a temperature of 2 MK, i.e. the same as that obtained from the partial-frame filter ratios. We therefore conclude that
leakage of visible light does not affect the partial frame Be-thin/Al-Poly
observations of this active region.

\begin{figure}[htbp!]
  \centering
  \includegraphics[width=.7\linewidth, angle=-90]{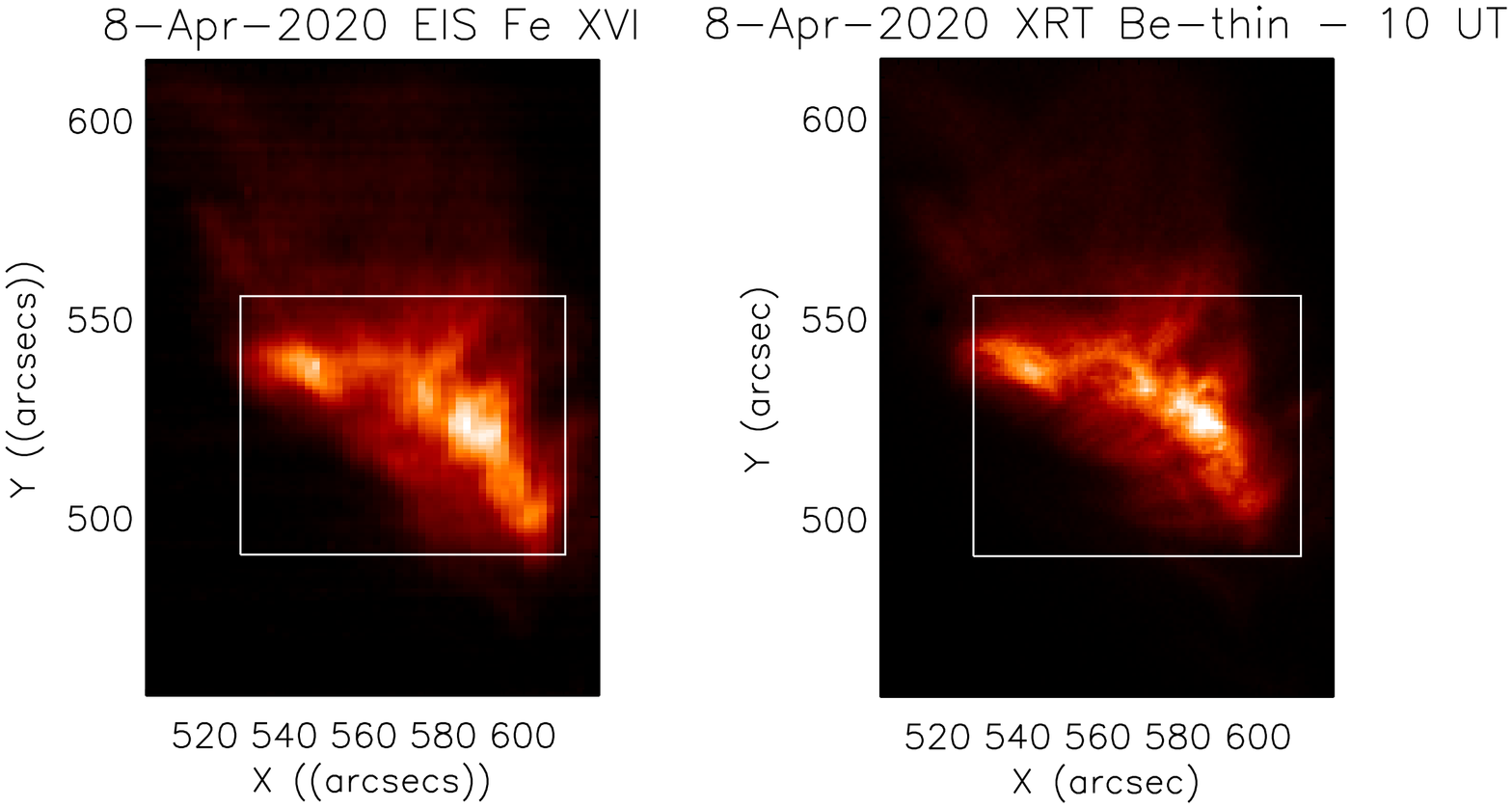} 
  \includegraphics[width=.5\linewidth, angle=-90]{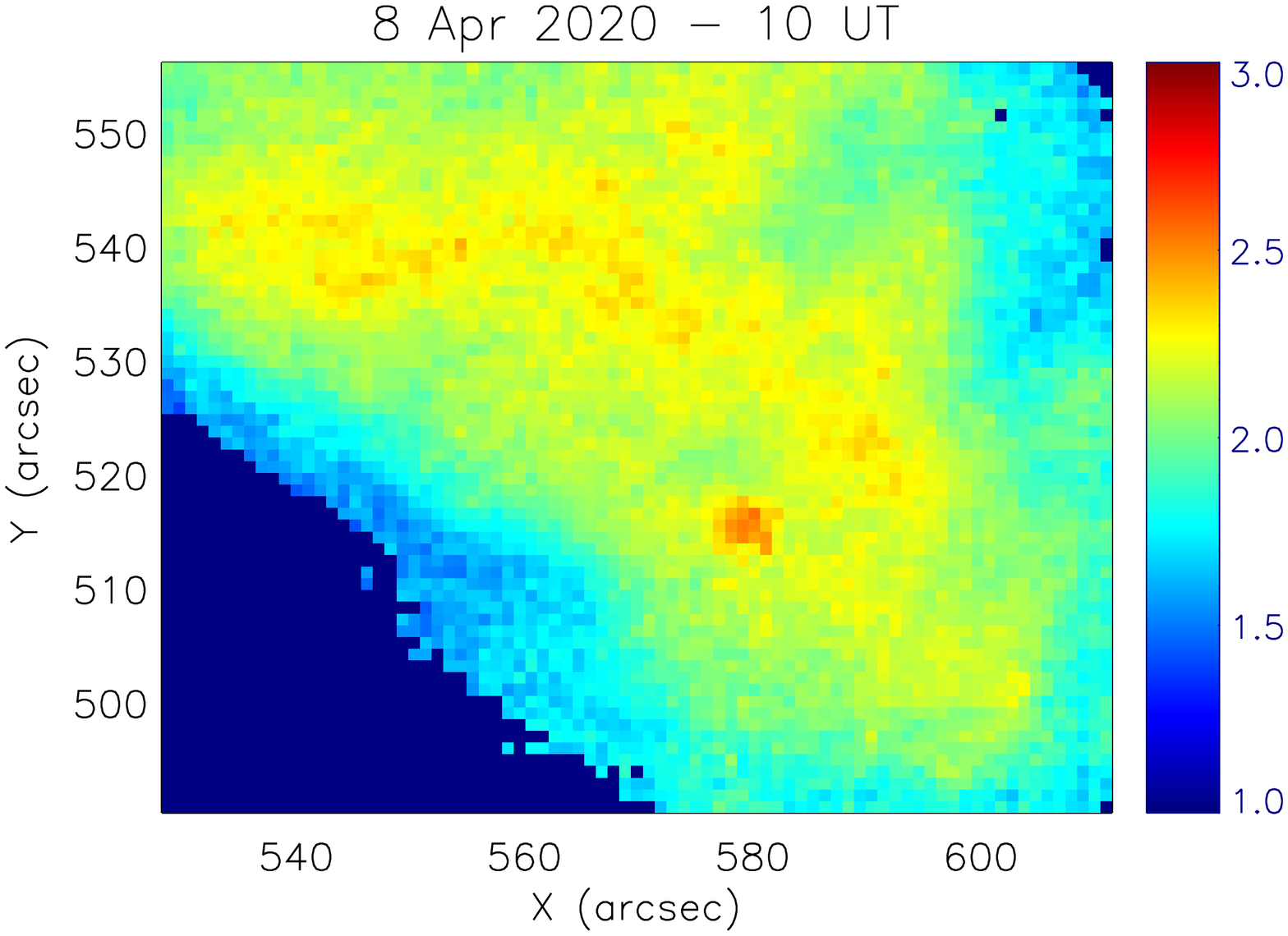}
  \caption{Top: the EIS raster image in \ion{Fe}{xvi} and an XRT Be-thin image
    of the AR core. The selected box was used to obtain a DEM from EIS and
    averaged signal in the XRT filters.
    Bottom: isothermal temperature (MK) map in the box region, obtained from the
 Be-thin/Al-Poly filter ratio. }
\label{fig:eis_xrt}
\end{figure}

To investigate this issue further, we analysed the XRT observations 
on April 8 and performed a cross-calibration with EIS.
A detailed pixel-by-pixel comparison is difficult, as the two
instruments have different spatial and temporal resolutions, and as
the EIS jitter is difficult to quantify. So we opted for an
overall cross-calibration of the whole AR core. 
We co-aligned the XRT images against EIS/AIA. Three Be-thin 
averaged frames around 10 UT are shown in Fig.~\ref{fig:eis_xrt} (top). We then
obtained from the  Be-thin/Al-Poly filter ratio 
the isothermal temperature  map in the AR core region (box),
shown in Fig.~\ref{fig:eis_xrt} (bottom). The averaged temperature is 2 MK.
The averaged XRT signal in the AR core in
each of the two filters changed by only a few percent during the EIS observation.
The averaged values are  31.6 DN/s in the Be-thin and 233.2 in the Al-Poly,
resulting in a ratio of 0.136. The ratio varied between 0.133 and 0.139
during the EIS scan of the AR core.

As we have a relatively reliable EIS absolute calibration, we can
predict the averaged AR core signal in the two XRT filters.
The main issue here is the chemical abundance of the elements. Having confirmed
from the EIS and XSM analysis that
the  Ar, S, Si, Fe abundances are consistent with the
\cite{delzanna:2013_multithermal} abundances, indicating an FIP effect of 3.2,
we have assumed that the abundances of the other elements contributing to
the XRT bands (mainly O, Ne) follow the same trend, which is
consistent to the X-ray  results obtained by \cite{delzanna_mason:2014}.

We used the DEM obtained from EIS, which has two peaks,
one around 1.8 MK and one around 2.5 MK, and calculated the XRT signal using the
current knowledge of the XRT effective areas as available in SolarSoft.
We obtained the simulated spectra shown in the Appendix and an averaged 
value of 25 DN/s in the Be-thin and 196 in the Al-Poly,
resulting in a ratio of 0.128, i.e. only 6\% lower than observed,
and  corresponding to an isothermal temperature of 2 MK.
The absolute DN/s are also very close, within 26 and 19\%, which is a
remarkable result, considering the uncertainty in the calibration of the instruments
and in the chemical abundances. 

In conclusion, the lower isothermal temperatures obtained
for the AR core by the XRT Be-thin/Al-Poly filter ratio are simply
due to the fact that the Al-Poly is sensitive to low temperatures,
so the isothermal temperature has a significant contribution 
from the peak in the emission measure around 1.8 MK.

\begin{figure*}[htbp!]
  \centering
  \includegraphics[width=1.\linewidth]{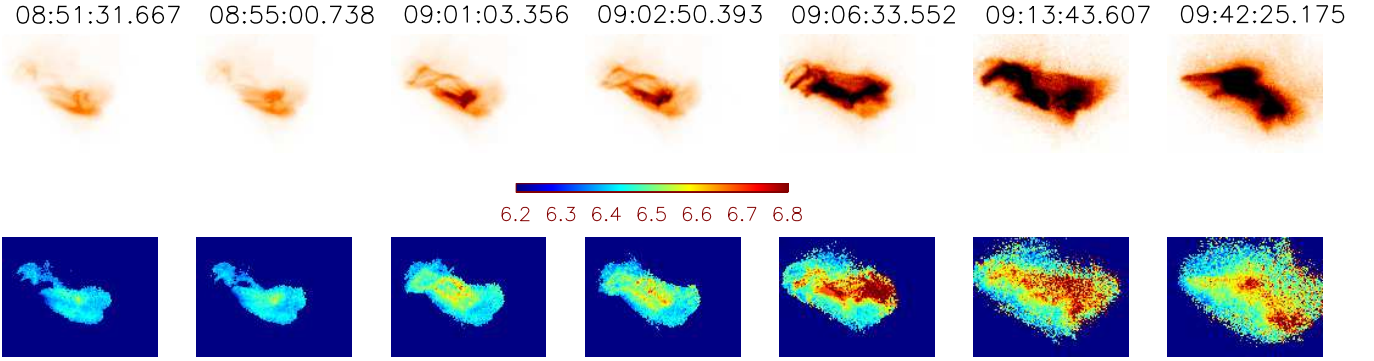}
  \caption{XRT negative images (top) and isothermal temperatures (bottom) 
  for a selection of times during the
  B-class flare on 2020 Apr 9.}
\label{fig:xrt_images}
\end{figure*}

\begin{figure}[htbp!]
  \centering
  \includegraphics[width=1.\linewidth]{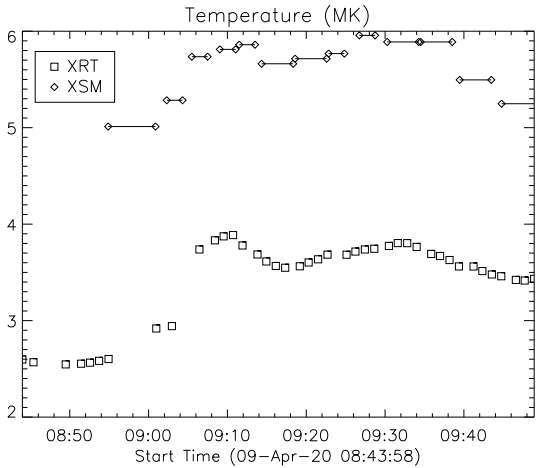}
  \caption{Isothermal temperatures obtained from XSM and XRT during the 
  Apr 9 9:30 UT flare.}
\label{fig:xrt_temps}
\end{figure}

\subsection{The 9:15 UT flare}

The AR produced a B-class flare which started around 9:00 UT and finished by 9:50 UT.
Fig.~\ref{fig:xrt_images} shows a selection of XRT images and
corresponding isothermal temperatures obtained from the Be-thin/Al-Poly filter ratio.
The two filters had exposures within a few seconds,  but several 
images were saturated and were discarded. 
The flare started with an activation of a filament and a compact X-ray emission.
Later on, many flare loops became activated and occupied a significant fraction
of the AR core, reaching  peak X-ray intensity around 9:15 UT.

From the XRT filter ratio we have also obtained the emission measure  for each
pixel. We have then obtained  averaged temperatures for the whole AR,
weighted by the EM values, to be compared to the isothermal temperatures obtained
from the spectral fitting of XSM. The XSM spectra were binned in time to increase the
signal to noise. 
The results are shown in Fig.~\ref{fig:xrt_temps}.
XSM indicates a peak temperature of nearly 6 MK, while the peak temperature
from XRT is much lower, although the variations are similar.
We have verified that the X-ray signal from the AR is dominating over 
that from the whole Sun, looking at the full-Sun X-ray Be-thin images.
Therefore, the XRT temperatures should be directly comparable to those of XSM.

 We note that 
fitting the XSM spectra with a two temperature component (one for the
quiescent AR core emission and one for the flare emission) 
would result in a slightly higher flare component, and the discrepancy 
would remain.

\begin{figure*}[ht!]
  \centering
  \includegraphics[width=.4\linewidth, angle=-270]{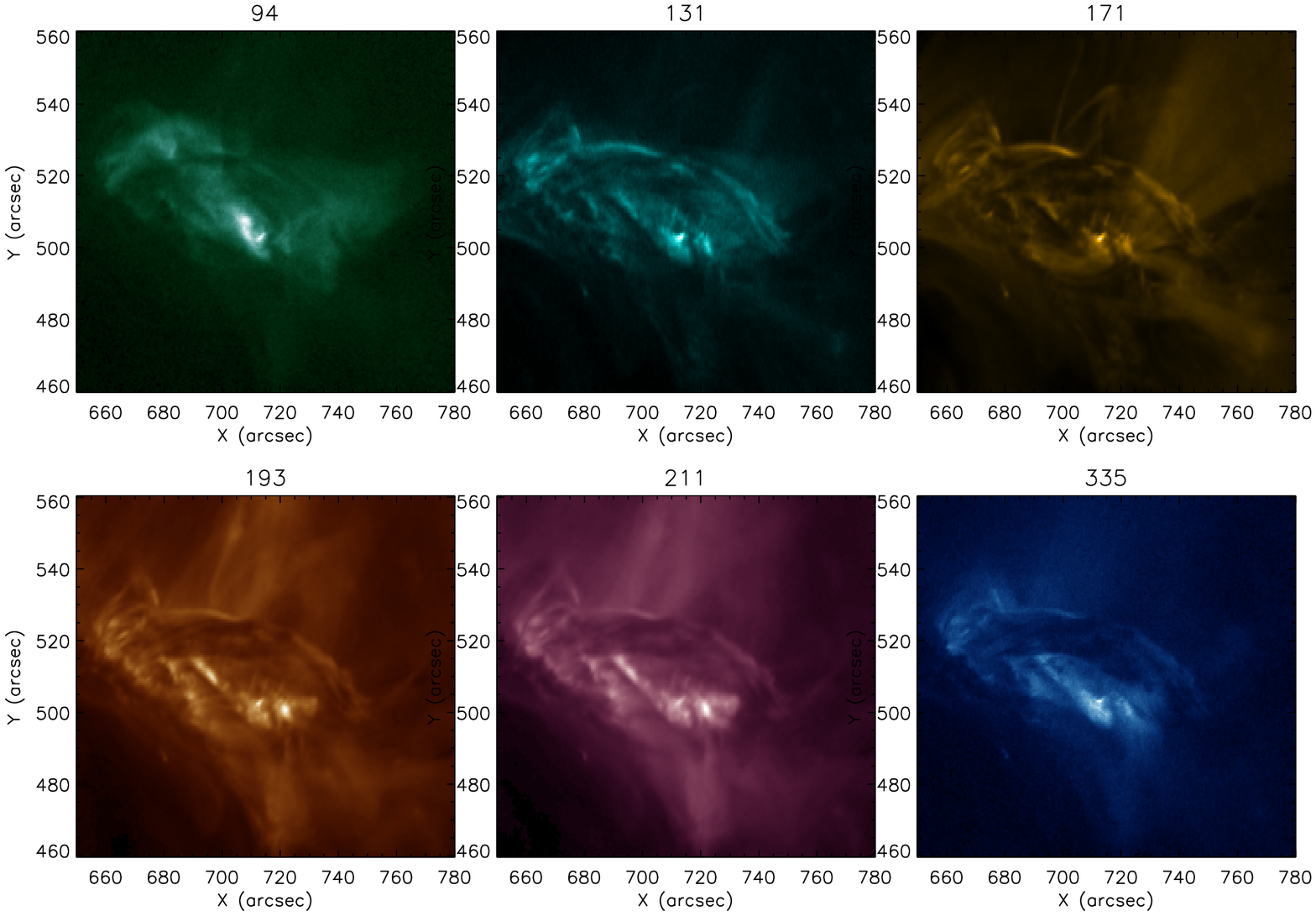} 
    \includegraphics[width=0.4\linewidth, angle=90]{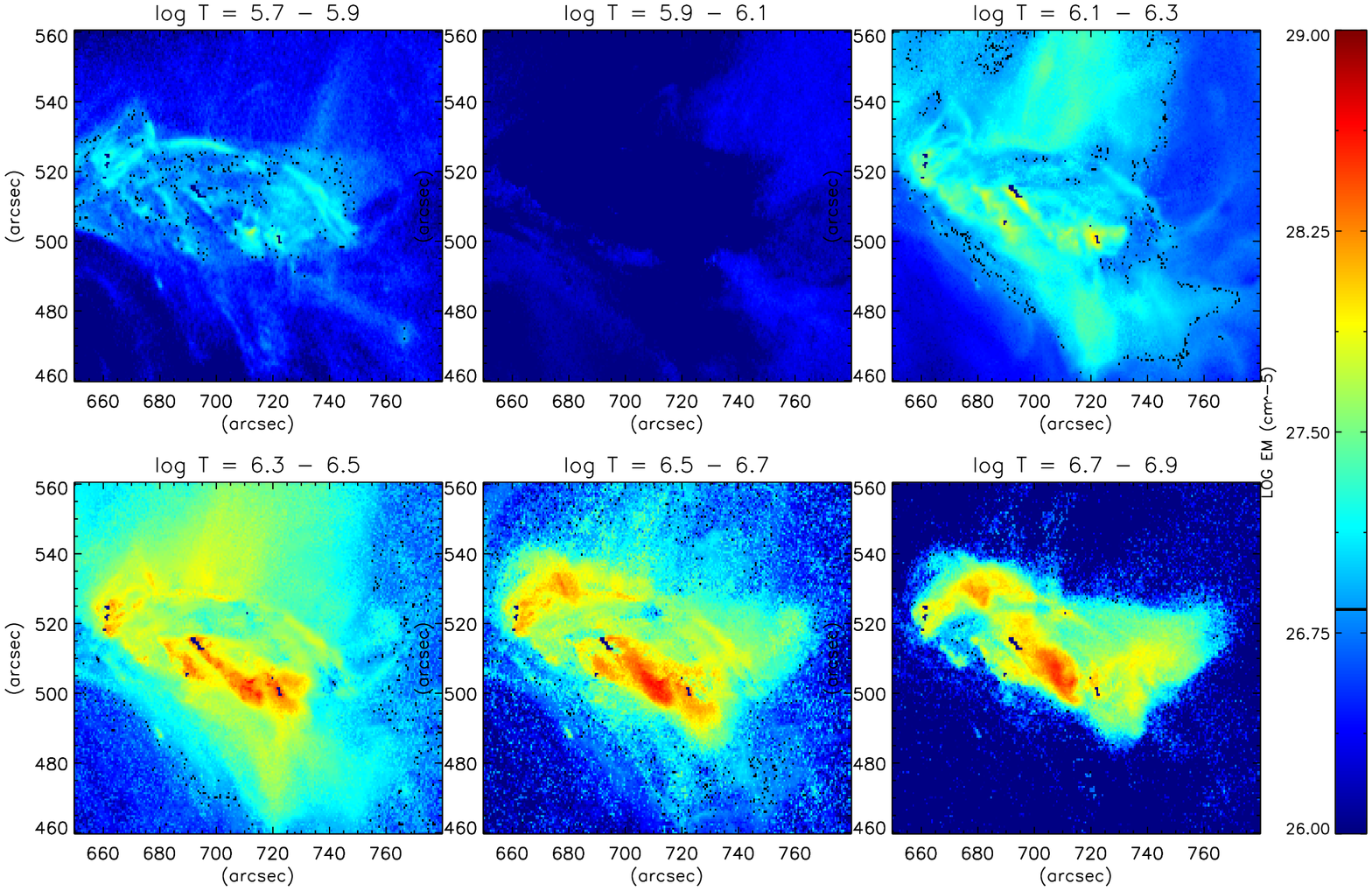} 
  \caption{Top: SDO AIA EUV images at 9:12 UT, during the peak of the microflare. 
  Bottom: DEM in selected temperature bins obtained from the images.}
\label{fig:aia_9:12}
\end{figure*}

As the isothermal assumption is clearly an approximation for both
the XSM and XRT datasets, we have explored the effects that a
DEM distribution has for this flare. We selected a time close to the peak,
at 9:12 UT, and obtained a DEM from the six AIA coronal images.
We averaged the AIA data over a minute around 9:12 UT.
Fig.~\ref{fig:aia_9:12} (top) shows the AIA EUV images.
Hot emission is visible in the AIA 94~\AA\ band, due to \ion{Fe}{xviii},
as described in \cite{delzanna:2013_multithermal}. Note that the filament is still
visible in absorption at this wavelengths, but not 
in the XRT X-ray images. It is clear from inspection of the
other AIA channels that this flare did not reach temperatures 
above 10 MK, otherwise the
\ion{Fe}{xxi} line would have contributed to the 131~\AA\ band.

\begin{figure}[ht!]
  \centering
  \includegraphics[clip,width=0.6\linewidth]{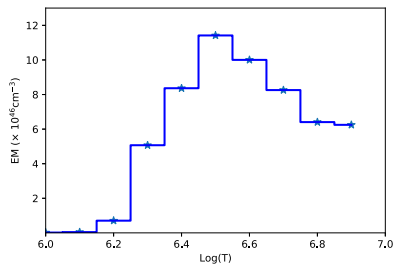}
  \caption{Averaged volume EM obtained from
   the AR core AIA EUV images at 9:12 UT. }
\label{fig:aia_demv}
\end{figure}

\begin{figure}[ht!]
  \centering
  \includegraphics[clip,width=0.7\linewidth]{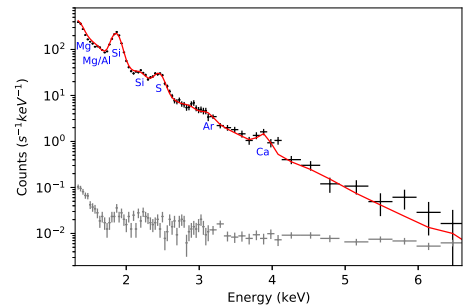}
  \caption{ XSM averaged spectrum at 9:12 UT, near the peak, with the
  predicted spectrum (red) based on the AIA EM.}
\label{fig:xsm_spectra}
\end{figure}

To obtain a DEM from the AIA data it is necessary to adopt a set of chemical
abundances. Note that the AIA bands are dominated by Fe lines, while
the XRT filters have contributions from Fe, Ne, O, Mg,  and other elements.
XSM can reliably measure the Si abundance, but also provides
abundances for other elements.
We performed an analysis of the XSM spectra around 9:12 UT and from the line
to continuum we obtained for 
Si, Mg, Al, and S, the following absolute abundances:
7.7, 7.5, 6.5, and 7.1 dex. The Mg, Al, and S are close to the photospheric
values, while the Si abundance is  between photospheric (7.51 dex) and the
\cite{delzanna:2013_multithermal} coronal value (8.0 dex).
We used these values for the DEM analysis.

For the other elements we have assumed the hybrid abundances tabulated in the CHIANTI file "sun\_coronal\_fludra\_1999\_ext.abund", 
as they provide  an intermediate FIP fractionation which is close to the 
results obtained from the low-FIP Si and the mid-FIP S, used as a proxy for the 
high-FIP elements. For example, the Fe hybrid
abundance is 7.83 dex (i.e. between the photospheric value of 7.50 and
the  coronal value of 8.0 dex, as Si), and the O, Ne hybrid 
abundances are actually close to the
photospheric values recommended by \cite{asplund_etal:2009}.

The resulting column EM values for a range of temperatures are shown in Fig.~\ref{fig:aia_9:12}  (bottom). 
We note that the high temperatures are not well constrained, as the only
hot channel with signal was the AIA 94~\AA. For this reason, the maximum temperature for the inversion was set
to log $T$[K]=6.9.

From the column EM values we  calculated a volume EM  
by averaging the emission measures over
the core of the microflare, in a  region covering 100 * 50 pixels in
the x and y direction, respectively.
The resulting  volume EM  is shown in Fig.~\ref{fig:aia_demv}.
We  then used this volume EM to predict the XSM spectrum, which is
shown, together with  the observed one, in Fig.~\ref{fig:xsm_spectra}.
There is excellent agreement between the two, indicating a reliable
cross-calibration between AIA and XSM.


We then preformed forward modelling of the XRT signal, using the same
EM and chemical abundances.
We obtained 222 and 1019 DN/s for the Be-thin and Al-Poly filters,
respectively. This  results  in a ratio of 0.22, equivalent to an 
isothermal temperature of 4 MK, nearly the same value obtained from the 
filter ratio.
We note that adopting a very different set of chemical abundances
does not change  the result. In fact,
with the photospheric values recommended by \cite{asplund_etal:2009}
or with the \cite{feldman:1992} coronal values we obtain a ratio of 0.23,
i.e. still 4 MK.


\section{Discussion and conclusions}

After a multi-wavelength survey of dozens of microflares and B-class flares
during the first period of operation of the XSM monitor, 
it became clear how difficult it is to obtain reliable measurements of these weak,
fast evolving events. 
XSM X-ray spectra have been great at providing an overview of how common 
microflares are in AR cores (even though  many small heating events also occur
outside ARs). Given its medium resolution, a multi-temperature analysis is 
difficult. However, the instrument has the great capability to measure 
the absolute chemical abundances for flares of B-class or above. 

Trying to catch a flare with a slit spectrometer such as EIS is notoriously difficult,
especially for  microflares, as they only last 10 minutes or so. 
The information that can be telemetered is also limited, so are the diagnostics.
On the basis of the present analysis, we developed
and tested new EIS observing sequences for additional diagnostics.

During the multi-wavelength campaign, we also learned how to adjust the
AEC to avoid saturation in the XRT filters as much as possible.
On the basis of the present analysis, we have run new XRT observing sequences
with different filters, as the Be-thin/Al-Poly provides isothermal temperatures
which are much lower than the peak values. 
With a direct comparison between simultaneous EIS and XRT observations we were able
to explain this result, due to the sensitivity of the Al-Poly filter to lower temperatures.
The improved XRT observations will be discussed in a following paper.

One important result of our analysis regards the 
quiescent AR core emission during its disk crossing, which was found to have
a distribution of temperatures and chemical abundances that did not change significantly over time. 
The XRT, EIS and XSM observations were all consistent with each other.

Most of the emission in the AR core was around 1.5 and 3 MK. 
As the XSM signal was dominated by the  AR core, 
we were also able to obtain another result: the XSM spectra during 
quiescence indicate a  FIP bias of about 2.5$\pm$0.2, adopting the  silicon value.
This result is important, as previous measurements 
did not clearly assess if it was the low-FIP elements that are enhanced in 
AR cores or the high-FIP ones to be depleted, relative to their photospheric
values. 
The EIS observations indicate that the Fe/Si vs. Ar/S abundances are increased by 
about a factor of about  3 over the photospheric value.
Combining the XSM and EIS results therefore indicates that the argon and sulphur 
abundances are slightly below their photospheric values. 
Such observations fit the nanoflare-heated scenario, where high-frequency 
nanoflares cause this relatively steady basal heating, and the Alfven waves excited by them
cause the chemical fractionation.
In fact, in the physical model discussed by \cite{laming_etal:2019},
a closed magnetic loop behaves as a resonant cavity, where  the waves 
bounce from footpoint to the other, with a travel time that is
an integral number of the wave half periods. In such case, 
the ponderomotive force creates a fractionation of the elements 
at the top of the chromosphere, where most of the  hydrogen is ionized,
and elements with a FIP around 10 eV such as sulphur are predicted to behave as
the high-FIP elements.

In our previous study \citep{mondal_etal:2021}, we used XSM data to 
show for the first time that chemical abundances of 
low-FIP elements  quickly
decrease towards photospheric values in B-class flares, with a return towards 
coronal values during the gradual phase.
AR 12759 produced many such flares during its disk crossing, but 
clearly such variations did not affect the abundances of the AR core.
In our previous paper, we have suggested possible reasons for such 
abundance variations, pointing out the need for a model to explain the observations.

It may well be that the same physical processes (e.g. magnetic reconnection 
in the corona) heat the basal AR cores and the flare loops, but produce very different signatures:
low-energy and high-frequency 
nanoflares would slowly evaporate already fractionated plasma from the top of the chromosphere, 
while B-class flares deposit their energy deeper in the chromosphere, causing a faster evaporation
of un-fractionated plasma.

The abundance variations during a flare could be used as a diagnostic to assess if the 
low-energy tail of the distribution of microflares is causing the basal active region heating.
Ideally, one would therefore want to see if also microflares have similar abundance 
variations as B-class flares. If they do, this would be an argument in favor of 
microflares not contributing to the basal AR core heating.

One of our aims was to provide estimates of the 
spatial and temperature distribution of the flare loops producing the XSM signal
using multi-wavelength observations, to complement the analysis of the
X-ray full-Sun spectra. The information on the spatial distribution is provided by the
AIA and XRT, but the temperature distribution turned out to be difficult to estimate.
The AIA analysis was limited, with only one hot band responding to the flare
(the 94~\AA). Assuming a continuous multi-thermal emission, we obtained from AIA 
a strong peak around  3 MK and a secondary (flare) emission around 6-7 MK.
The 3 MK emission is mostly background/foreground quiescent emission of the AR core.
We are encouraged by the excellent agreement between the 
X-ray XSM spectra as predicted from the AIA DEM modelling and the observed values,
especially considering the uncertainty in the AIA calibration. 
However, the isothermal approximation fits the XSM spectra equally well to
a multi-thermal distribution. 
Clearly, the isothermal temperatures from the XSM analysis are more representative
of the `true' peak flare temperatures than those from 
the XRT Be-thin/Al-Poly  filter ratio, as the latter has a bias towards the lower
temperatures. 
However, multi-thermal emission is clearly present and should be taken into account 
whenever suitable observations are available.

Better estimates of the temperature distribution can be obtained from 
spectroscopic observations or from AIA when 
temperatures  well above 7 MK are present, as Fe XXI becomes visible 
in the AIA 131~\AA\ (and also Fe XXIV in the 193~\AA\ band). 
Also, the XSM spectra of higher-T flares are very different, not just in terms
of slope of the continuum, but also in terms of line emission.

In some respects we were surprised to see a low maximum temperature of about 6-7 MK
from this B-class flare, as smaller flares often do produce much higher
temperatures, as in the \cite{delzanna_etal:2011_flare} example. 
In this case it is clear that the higher GOES class is simply due to 
the presence of an extended set of flare loops, which fits with the conclusions 
on the C-class flares 
by \citet{Bowen2013} that flares with larger volumes have lower maximum temperatures.

Further improvements on our understanding of small flares is achievable 
with current instrumentation and modelling, but as we mentioned in the 
introduction, key new high-resolution spectroscopic observations are needed.

\acknowledgments
 GDZ and HEM acknowledge support from STFC (UK) via the consolidated grants 
 to the atomic astrophysics group at DAMTP, University of Cambridge (ST/P000665/1. and ST/T000481/1).  KKR acknowledges support from the NASA HSO Connect program, grant number 80NSSC20K1283.
 AB was a J.C. Bose National Fellow 
during the period of this work.
We acknowledge the use of data from the Solar X-ray Monitor (XSM) on board the Chandrayaan-2 mission of the Indian Space Research Organisation (ISRO), archived at the Indian Space Science Data Centre (ISSDC).XSM was developed by Physical Research Laboratory (PRL) with support from various ISRO centers. Research at PRL is supported by the Department
of Space, Govt. of India. The collaboration between the PRL and Cambridge groups 
has been facilitated through the Royal Society Grant No. IES-R2-170199. Hinode is a Japanese mission developed and launched by ISAS/ JAXA, with NAOJ as domestic partner and NASA and STFC (UK) as international partners. It is operated by these agencies in cooperation with ESA and NSC (Norway). AIA data are courtesy of SDO (NASA) and the AIA
consortium.
 CHIANTI is a collaborative project involving the University of Cambridge (UK),
George Mason University, and the University of Michigan (USA).
We thank the referee for useful comments on the manuscript.
 
 
 \bibliography{paper}{}

\bibliographystyle{aasjournal}



 \appendix

\section{Further details}

\begin{figure}[ht!]
\begin{center}
  \includegraphics[width=0.4\linewidth, angle=-90]{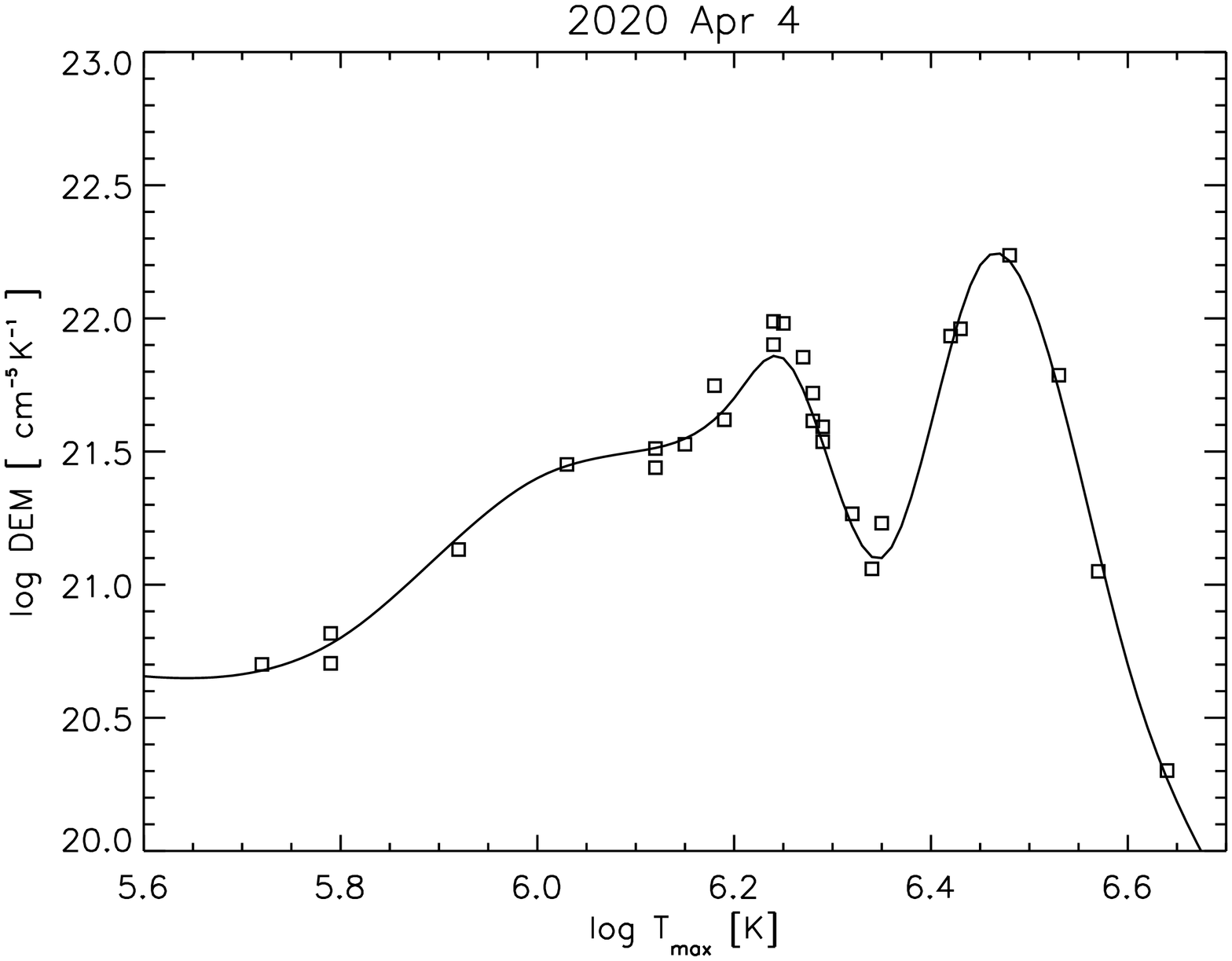}
  \includegraphics[width=0.4\linewidth, angle=-90]{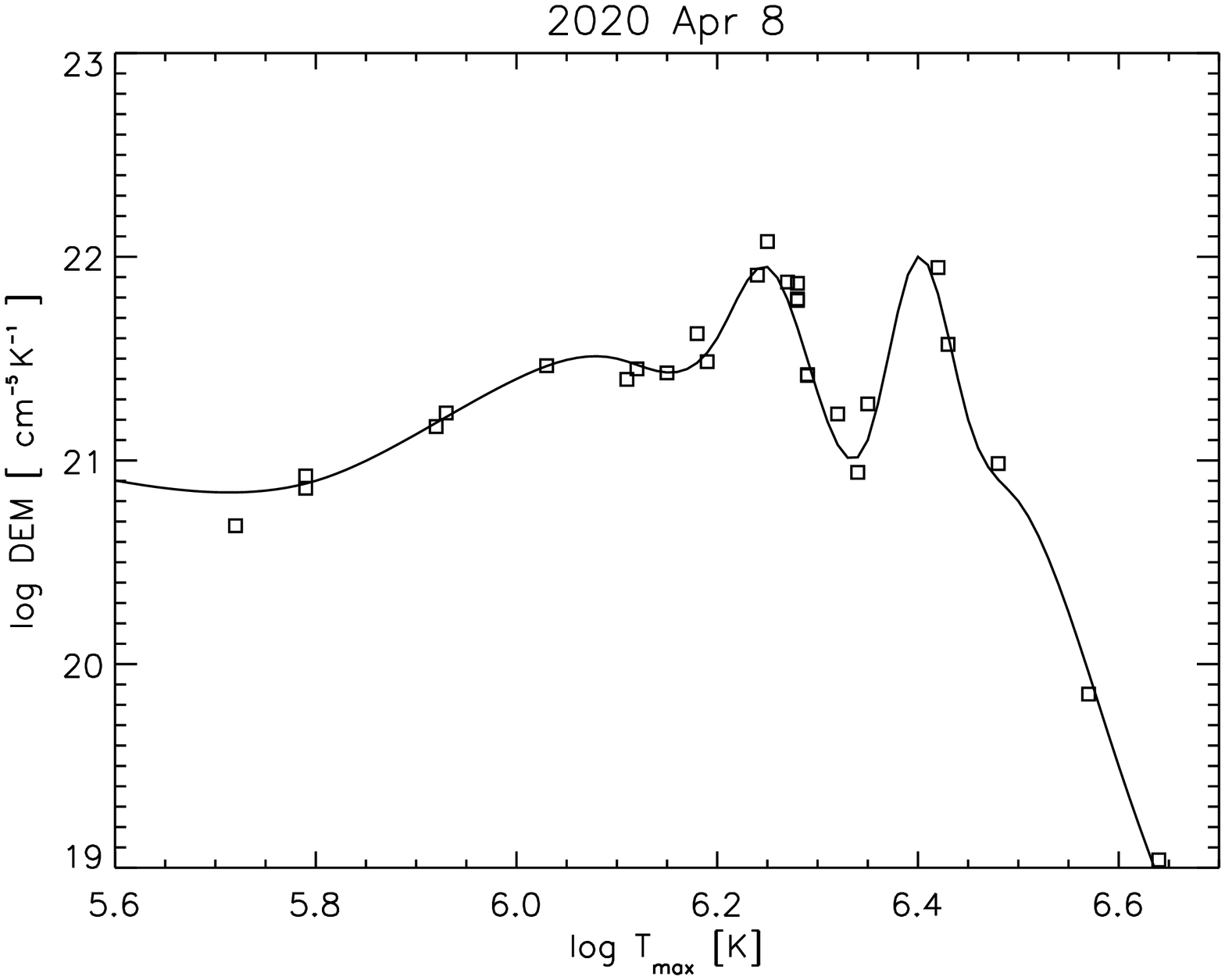}
  \end{center}
  \caption{DEM for the AR core observed on 2020 Apr 4 and 8 obtained from Hinode EIS.
  The points are plotted at the temperature $T_{\rm max}$, 
 and at the theoretical vs. the observed
intensity ratio multiplied by the DEM value.}
\label{fig:eis_dem2}
\end{figure}

Fig.~\ref{fig:eis_dem2} shows the  two DEMs for the AR core observed on 2020 Apr 4 and 8 obtained from Hinode EIS, while Tables~\ref{tab:list2},\ref{tab:list3} provide the details.

\def\baselinestretch{1.}
\begin{table}[!htbp]
  \caption{List of  emission lines from the 2020 Apr 4 active region core.
  $\lambda_{\rm obs}$ (\AA) is the measured wavelength.
DN is the number of total counts in the line, while $I_{\rm obs}$
is the calibrated radiance (phot cm$^{-2}$ s$^{-1}$ 
arc-second$^{-2}$) obtained with our new EIS calibration.
$T_{\rm max}$ and $T_{\rm eff}$ are the maximum and effective temperature (log values in K),
$R$ the ratio between the predicted and observed radiances, 
Ion and $\lambda_{\rm exp}$ (\AA) the main contributing line, and 
$r$ the fractional contribution to the blend.}
\begin{center} 
\scriptsize
\begin{tabular}{@{}lllllllll@{}}
 \hline\hline \noalign{\smallskip}
  $\lambda_{\rm obs}$  & DN & $I_{\rm obs}$   &  $T_{\rm max}$ & $T_{\rm eff}$  & $R$ & Ion & $\lambda_{\rm exp}$   &  $r$ \\
 \hline \noalign{\smallskip}
 
 195.97 &      810 & 12.4 &  5.72 &  5.89 &  0.95 &  \ion{Fe}{viii} &  195.972 & 0.83 \\ 
  
 280.73 &       55 & 15.5 &  5.79 &  5.93 &  0.91 &  \ion{Mg}{vii} &  280.742 & 0.93 \\ 
 
 275.37 &      300 & 43.2 &  5.79 &  5.93 &  1.18 &  \ion{Si}{vii} &  275.361 & 0.98 \\ 
 
 197.85 &     1260 & 23.3 &  5.93 &  6.03 &  1.00 &  \ion{Fe}{ix} &  197.854 & 0.90 \\ 
 
 188.49 &     1650 & 52.6 &  5.92 &  6.04 &  1.11 &  \ion{Fe}{ix} &  188.493 & 0.80 \\

 184.53 &     2120 & 158.0 &  6.03 &  6.10 &  0.99 &  \ion{Fe}{x} &  184.537 & 0.95 \\

 192.80 &     4920 & 95.4 &  6.12 &  6.16 &  1.00 &  \ion{Fe}{xi} &  192.813 & 0.95 \\ 
  
 180.40 &     1910 & 703.0 &  6.12 &  6.16 &  1.19 &  \ion{Fe}{xi} &  180.401 & 0.97 \\

 271.99 &      798 & 90.9 &  6.15 &  6.19 &  1.05 &  \ion{Si}{x} &  271.992 & 0.97 \\

 192.39 &    10400 & 215.0 &  6.19 &  6.21 &  1.09 &  \ion{Fe}{xii} &  192.394 & 0.96 \\

 264.23 &      606 & 70.9 &  6.18 &  6.23 &  0.75 &  \ion{S}{x} &  264.230 & 0.98 \\

 251.95 &     1200 & 304.0 &  6.24 &  6.27 &  0.91 &  \ion{Fe}{xiii} &  251.952 & 0.97 \\ 
 
 202.04 &     8150 & 460.0 &  6.25 &  6.27 &  0.74 &  \ion{Fe}{xiii} &  202.044 & 0.96 \\

 188.67 &     1010 & 31.1 &  6.24 &  6.28 &  0.74 &  \ion{S}{xi} &  188.675 & 0.62 \\ 
                                 &   &  &  &  &  \ion{Fe}{xi} &  188.630 & 0.16 \\ 
                                 &   &  &  &  &  \ion{Fe}{ix} &  188.681 & 0.10 \\ 
 
 281.42 &       60 & 18.1 &  6.28 &  6.32 &  1.05 &  \ion{S}{xi} &  281.402 & 0.96 \\ 
 
 285.83 &       67 & 37.5 &  6.27 &  6.32 &  0.76 &  \ion{S}{xi} &  285.822 & 0.97 \\ 
 
 
 246.89 &       67 & 27.2 &  6.28 &  6.32 &  0.83 &  \ion{S}{xi} &  246.895 & 0.97 \\

 264.79 &     5380 & 626.0 &  6.29 &  6.34 &  0.87 &  \ion{Fe}{xiv} &  264.788 & 0.93 \\

 
 211.31 &     2190 & 751.0 &  6.29 &  6.35 &  0.99 &  \ion{Fe}{xiv} &  211.317 & 0.98 \\

 278.27 &       21 & 4.7 &  6.32 &  6.39 &  0.90 &  \ion{P}{xii} &  278.286 & 0.94 \\ 
 
 288.40 &      105 & 89.8 &  6.35 &  6.41 &  0.74 &  \ion{S}{xii} &  288.434 & 0.98 \\

 
 284.16 &    10100 & 4350.0 &  6.34 &  6.43 &  1.11 &  \ion{Fe}{xv} &  284.163 & 0.99 \\ 
 
 194.41 &      553 & 9.3 &  6.53 &  6.44 &  0.88 &  \ion{Ar}{xiv} &  194.401 & 0.80 \\

 183.45 &       87 & 9.1 &  5.35 &  6.45 &  1.18 &  \ion{Ca}{xiv} &  183.460 & 0.85 \\ 
 
 256.68 &     2080 & 339.0 &  6.42 &  6.46 &  0.90 &  \ion{S}{xiii} &  256.685 & 0.98 \\ 
 
 
 
 262.98 &     2160 & 258.0 &  6.43 &  6.47 &  1.14 &  \ion{Fe}{xvi} &  262.976 & 0.97 \\ 
 
 249.18 &      601 & 201.0 &  6.48 &  6.48 &  0.95 &  \ion{Ni}{xvii} &  249.189 & 0.97 \\ 
 
 200.99 &      324 & 11.8 &  6.64 &  6.48 &  0.93 &  \ion{Ca}{xv} &  200.977 & 0.86 \\ 
 
 193.87 &     1340 & 23.7 &  6.57 &  6.49 &  1.21 &  \ion{Ca}{xiv} &  193.866 & 0.96 \\

\noalign{\smallskip}\hline                                   
\end{tabular}
\normalsize
\end{center}
\label{tab:list2}
\end{table}

\def\baselinestretch{1.}
\begin{table}[!htbp]
  \caption{List of  emission lines from the 2020 Apr 8 active region core.
  $\lambda_{\rm obs}$ (\AA) is the measured wavelength.
DN is the number of total counts in the line, while $I_{\rm obs}$
is the calibrated radiance (phot cm$^{-2}$ s$^{-1}$ 
arc-second$^{-2}$) obtained with our new EIS calibration.
$T_{\rm max}$ and $T_{\rm eff}$ are the maximum and effective temperature (log values in K),
$R$ the ratio between the predicted and observed radiances, 
Ion and $\lambda_{\rm exp}$ (\AA) the main contributing line, and 
$r$ the fractional contribution to the blend.}
\begin{center} 
\scriptsize
\begin{tabular}{@{}lllllllll@{}}
 \hline\hline \noalign{\smallskip}
  $\lambda_{\rm obs}$  & DN & $I_{\rm obs}$   &  $T_{\rm max}$ & $T_{\rm eff}$  & $R$ & Ion & $\lambda_{\rm exp}$   &  $r$ \\
 \hline \noalign{\smallskip}

 195.98 &      676 & 10.4 &  5.72 &  5.83 &  1.46 &  \ion{O}{iv} &  196.006 & 0.15 \\ 
                                 &   &  &  &  &  \ion{Fe}{viii} &  195.972 & 0.78 \\ 
 
 280.73 &       62 & 17.5 &  5.79 &  5.90 &  0.92 &  \ion{Mg}{vii} &  280.742 & 0.95 \\ 
  
 275.36 &      375 & 54.1 &  5.79 &  5.91 &  1.05 &  \ion{Si}{vii} &  275.361 & 0.98 \\ 
 
 188.50 &     1620 & 51.7 &  5.92 &  6.01 &  1.04 &  \ion{Fe}{ix} &  188.493 & 0.89 \\ 
  
 197.87 &     1360 & 25.0 &  5.93 &  6.03 &  0.95 &  \ion{Fe}{ix} &  197.854 & 0.90 \\

 184.54 &     2050 & 152.0 &  6.03 &  6.10 &  1.00 &  \ion{Fe}{x} &  184.537 & 0.95 \\ 
 
 192.82 &     4580 & 88.8 &  6.12 &  6.15 &  1.05 &  \ion{Fe}{xi} &  192.813 & 0.93 \\ 
 
 180.41 &     1890 & 695.0 &  6.11 &  6.16 &  1.22 &  \ion{Fe}{xi} &  180.401 & 0.91 \\

 271.98 &      795 & 90.6 &  6.15 &  6.19 &  1.00 &  \ion{Si}{x} &  271.992 & 0.96 \\ 
  
 192.40 &    10000 & 207.0 &  6.19 &  6.22 &  1.11 &  \ion{Fe}{xii} &  192.394 & 0.95 \\

 264.23 &      606 & 70.9 &  6.18 &  6.23 &  0.72 &  \ion{S}{x} &  264.230 & 0.98 \\

 251.94 &      994 & 252.0 &  6.24 &  6.26 &  1.08 &  \ion{Fe}{xiii} &  251.952 & 0.97 \\ 
 
202.05 &     7900 & 446.0 &  6.25 &  6.26 &  0.75 &  \ion{Fe}{xiii} &  202.044 & 0.96 \\ 
 
 188.68 &      773 & 23.7 &  6.28 &  6.26 &  0.72 &  \ion{S}{xi} &  188.675 & 0.74 \\ 
                                 &   &  &  &  &  \ion{Fe}{ix} &  188.681 & 0.14 \\

 281.41 &       78 & 23.4 &  6.28 &  6.29 &  0.73 &  \ion{S}{xi} &  281.402 & 0.95 \\ 
 
 285.83 &       55 & 30.9 &  6.27 &  6.29 &  0.82 &  \ion{S}{xi} &  285.822 & 0.96 \\ 
  
 246.90 &       82 & 33.4 &  6.28 &  6.29 &  0.60 &  \ion{S}{xi} &  246.895 & 0.97 \\

 264.78 &     3370 & 393.0 &  6.29 &  6.30 &  1.19 &  \ion{Fe}{xiv} &  264.788 & 0.92 \\ 
  
 211.32 &     1560 & 534.0 &  6.29 &  6.31 &  1.18 &  \ion{Fe}{xiv} &  211.317 & 0.97 \\ 
 
 194.40 &     1000 & 1.7 &  6.53 &  6.31 &  1.35 &  \ion{Ar}{xiv} &  194.401 & 0.49 \\ 
                                 &   &  &  &  &  \ion{Fe}{xii} &  194.377 & 0.10 \\ 
                                 &   &  &  &  &  \ion{Mn}{x} &  194.364 & 0.18 \\ 
 
 278.26 &       19 & 4.3 &  6.32 &  6.31 &  0.71 &  \ion{P}{xii} &  278.286 & 0.89 \\

 201.00 &       66 & 2.4 &  6.64 &  6.34 &  0.80 &  \ion{Ca}{xv} &  200.977 & 0.50 \\ 
                                 &   &  &  &  &  \ion{Fe}{xii} &  200.978 & 0.12 \\ 
  
 288.40 &       75 & 64.1 &  6.35 &  6.35 &  0.66 &  \ion{S}{xii} &  288.434 & 0.97 \\ 
 
 284.15 &     5300 & 2280.0 &  6.34 &  6.36 &  1.19 &  \ion{Fe}{xv} &  284.163 & 0.98 \\ 
 
 256.68 &      961 & 157.0 &  6.42 &  6.40 &  0.74 &  \ion{S}{xiii} &  256.685 & 0.97 \\ 
 
 262.98 &      713 & 85.3 &  6.43 &  6.40 &  1.11 &  \ion{Fe}{xvi} &  262.976 & 0.98 \\ 
 
 193.87 &      224 & 4.0 &  6.57 &  6.41 &  1.28 &  \ion{Ca}{xiv} &  193.866 & 0.90 \\ 
 
 249.17 &      147 & 49.1 &  6.48 &  6.43 &  0.83 &  \ion{Ni}{xvii} &  249.189 & 0.95 \\ 

\noalign{\smallskip}\hline                                   
\end{tabular}
\normalsize
\end{center}
\label{tab:list3}
\end{table}

\section{EIS calibration}

A DEM analysis was applied to
off-limb quiet Sun observations close in time to the
observations discussed here, to obtain  the relative EIS calibration
using the strongest coronal lines. The advantage is that the
plasma is nearly isothermal and isodensity, and 
removes from the coronal lines blending with cool lines. 
This is an extension of the method used by \cite{warren_etal:2014},
where strict isothermality was assumed. 
The line ratio method adopted in \cite{delzanna:13_eis_calib}
was also extended, using both  quiet Sun and Active region 
observations. 
The calibration relative to AIA was also assessed,
as was AIA against SDO EVE. Further details are given in 
Del Zanna and Warren (2022).


\section{XRT forward modelling}

\begin{figure}[ht!]
\begin{center}
  \includegraphics[width=0.6\linewidth, angle=0]{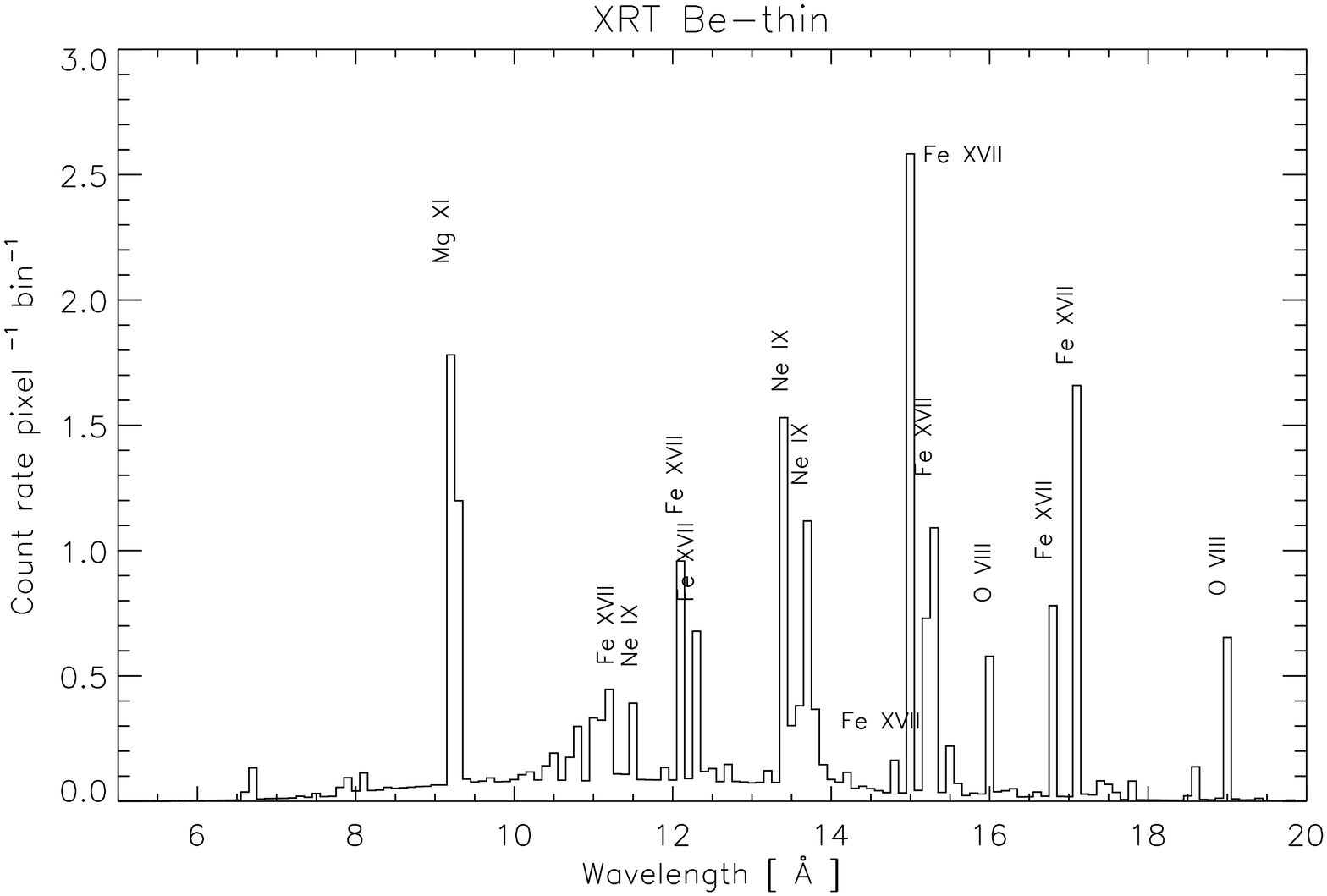}
  \includegraphics[width=0.6\linewidth, angle=0]{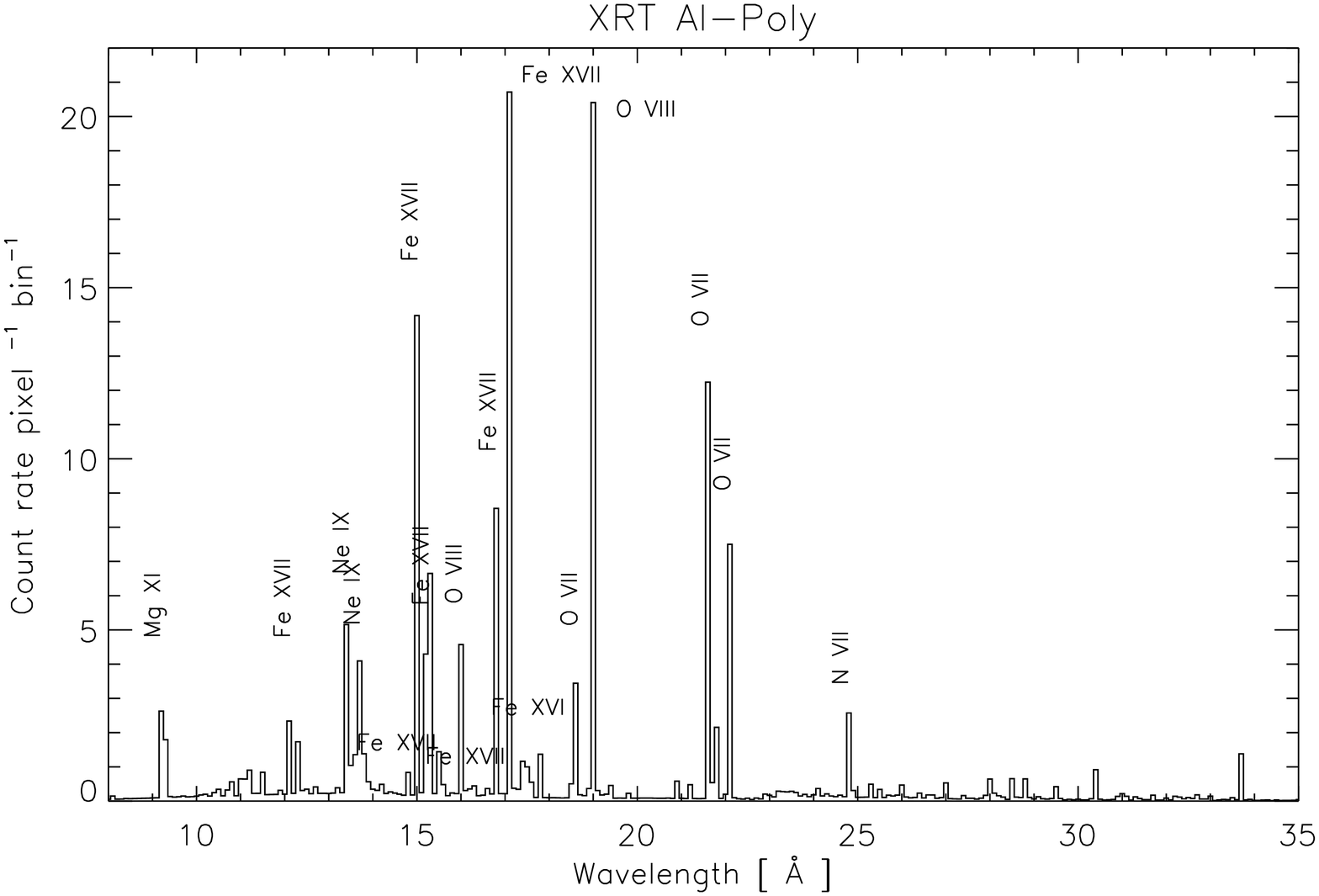}
\end{center}
  \caption{Simulated XRT signal for the quiescent AR core on Apr 8.}
\label{fig:xrt_arcore}
\end{figure}

Fig.~\ref{fig:xrt_arcore}  shows the simulated XRT signal 
for the quiescent AR core on Apr 8.

\end{document}